\documentclass[nofootinbib,twocolumn,prd,groupedaddress,preprintnumbers]{revtex4-1}
\pdfoutput=1
\usepackage[utf8]{inputenc}
\usepackage{slashed}
\usepackage{bm}
\usepackage{graphicx}
\usepackage[caption=false]{subfig}
\usepackage{color}
\usepackage{upgreek}
\usepackage{amssymb}
\usepackage{mathrsfs}
\usepackage{amsmath}
\usepackage[normalem]{ulem}
\usepackage{bbm}
\usepackage{array}
\usepackage[usenames,dvipsnames]{xcolor}
\RequirePackage[colorlinks=true,urlcolor=Magenta,anchorcolor=blue,citecolor=blue,filecolor=blue,
linkcolor=Magenta,menucolor=blue,linktocpage=true,pdfproducer=medialab]{hyperref}
\usepackage{catchfile} 
\newcommand{\getenv}[2][]{%
  \CatchFileEdef{\temp}{"|kpsewhich --var-value #2"}{}%
  \if\relax\detokenize{#1}\relax\temp\else\let#1\temp\fi}
\getenv[\USER]{USER}
\def\diag{\mathtt{diag}}
\def\tr{{\rm Tr}}
\def\hc{\text{h.c.}}
\def\Im{\text{Im}}

\def\unity{\mathbbm{1}}

\newcommand{\ov}[1]{\overline{#1}}
\def\be{\begin{equation}}
\def\ee{\end{equation}}

\newcommand{\GeV}{\;\text{GeV}}
\newcommand{\MeV}{\;\text{MeV}}
\newcommand{\eV}{\;\text{eV}}
\newcommand{\Dmsol}{\Delta m^2_\text{sol}}
\newcommand{\Dmatm}{\Delta m^2_\text{atm}}
\newcommand{\ml}{m_\text{light}}
\newcommand{\Ml}{M_\text{light}}

\def\d{\mathrm{d}}
\def\D{\mathscr{D}}
\def\G{\mathcal{G}}
\def\I{\mathcal{I}}
\def\K{\mathscr{K}}
\def\L{\mathscr{L}}
\def\O{\mathcal{O}}
\def\P{\mathcal{P}}
\def\sP{\mathscr{P}}
\def\sR{\mathscr{R}}
\def\W{\mathscr{W}}
\newcommand{\blue}[1]{\color{blue} #1 \color{black} }

\begin{document}

\title{\blue{Predictive Leptogenesis from Minimal Lepton Flavour Violation}}

\author{{\bf L. Merlo}}
\email{luca.merlo@uam.es}
\author{{\bf S. Rosauro-Alcaraz}}
\email{salvador.rosauro@uam.es}
\affiliation{\vspace{1mm} 
Departamento de F\'isica Te\'orica and Instituto de F\'isica Te\'orica, IFT-UAM/CSIC, Universidad Aut\'onoma de Madrid, Cantoblanco, 28049, Madrid, Spain}

\begin{abstract}
A predictive Leptogenesis scenario is presented based on the Minimal Lepton Flavour Violation symmetry. In the realisation with three right-handed neutrinos transforming under the same flavour symmetry of the lepton electroweak doublets, lepton masses and PMNS mixing parameters can be described according to the current data, including a large Dirac CP phase. The observed matter-antimatter asymmetry of the Universe can be achieved through Leptogenesis, with the CP asymmetry parameter $\varepsilon$ described in terms of only lepton masses, mixings and phases, plus two real parameters of the low-energy effective description. This is in contrast with the large majority of models present in the literature, where $\varepsilon$ depends on several high-energy parameters, preventing a direct connection between low-energy observables and the baryon to photon ratio today. Recovering the correct amount of baryon asymmetry in the Universe constrains the Majorana phases of the PMNS matrix within specific ranges of values: clear predictions for the neutrinoless double beta decay emerge, representing a potential smoking gun for this framework.
\end{abstract}

\preprint{FTUAM-18-1}
\preprint{IFT-UAM/CSIC-18-004}

\maketitle

%
%
\section{Introduction}

The uncertainties on the measurements for the cosmic abundances of the lightest elements have improved considerably in the last decades, posing stringent constraints on the thermal history of the very early Universe. The observed abundances of protium, deuterium, ${}^3$He, ${}^4$He and Lithium, besides well agreeing with the predictions of the standard Big Bang Nucleosynthesis~\cite{Wagoner:1972jh}, allow to deduce the value of the baryon to photon ratio today,
\be
\eta_B\equiv\dfrac{N_B-N_{\bar{B}}}{N_\gamma}\,,
\ee
where $N_{B,\bar{B},\gamma}$ are the number densities of baryons, anti-baryons and photons, respectively. An independent determination of $\eta_B$ is provided by the CMB measurements ~\cite{Ade:2015xua} that agrees with the value extracted from the lightest element abundances: 
\be
\eta_B=(6.11 \pm 0.04) \times 10^{-10}\,.
\label{ExpBAU}
\ee
Despite being so tiny, this non-vanishing value poses one of the most relevant unsolved questions in particle physics and cosmology today: why are there more baryons than anti-baryons in the present Universe?

In 1967, Sakharov first suggested that the baryon asymmetry of the Universe (BAU) might not represent some sort of initial condition, but could be understood in terms of microphysical laws that fulfil the following three conditions~\cite{Sakharov:1967dj}:
\begin{itemize}
\item[-] Baryon number violation
\item[-] C and CP violation
\item[-] Departure from Thermal Equilibrium.
\end{itemize}
Although Standard Model (SM) interactions satisfy all these three requirements, there is not enough CP violation to produce the measured value of BAU~\cite{Jarlskog:1985ht,Kuzmin:1985mm}. Several alternative mechanisms have been proposed, typically extending the SM spectrum and/or its symmetries: GUT Baryogenesis, Electroweak (EW) Baryogenesis, the Affleck-Dine mechanism, Leptogenesis. The latter will be the focus in this paper as Baryogenesis through Leptogenesis~\cite{Fukugita:1986hr} (see Refs.~\cite{Dev:2017trv,Drewes:2017zyw,Dev:2017wwc,Biondini:2017rpb,Chun:2017spz,Hagedorn:2017wjy} for update reviews on the subject), besides being promising in a large part of the associated parameter space, represents a framework where also other open problems of the SM of particle physics may find a solution: on the one hand the origin of neutrino masses and on the other hand the Flavour problem. 

The small, but non-vanishing, masses of the light active neutrinos represent an experimental evidence of the incompleteness of the SM. The introduction of right-handed (RH) neutrinos {\it \`a la} type I Seesaw mechanism~\cite{Minkowski:1977sc,GellMann:1980vs,Yanagida:1980xy,Schechter:1980gr,Mohapatra:1980yp} is an elegant approach that explains the smallness of the active neutrino masses through the largeness of the masses of their RH counterparts. This mechanism provides the ingredients to explain the present amount of BAU: there is a leptonic source of CP violation and a source of Lepton number violation; RH neutrino decays may occur out-of-equilibrium, when the temperature drops below their masses. In consequence, out-of-equilibrium decays of the RH neutrinos might produce a lepton asymmetry that is then partially converted into BAU through non-perturbative sphaleron effects~\cite{Buchmuller:2005eh}: in the SM context,
\be
B=-\dfrac{c_s}{1-c_s}L\qquad\text{with}\qquad
c_s\equiv\dfrac{8N_F+4}{22N_F+13}<1\,,
\label{cs}
\ee
with $N_F$ the number of flavour species considered. It follows that the more anti-leptons are produced, the more baryons are generated, with a rate that is approximately close to $1/3$. 

The basic quantity in Leptogenesis is the parameter $\varepsilon$ that measures the amount of CP asymmetry generated in the decays of the RH neutrinos $\nu_R$~\cite{Fukugita:1986hr}: indicating with $\Gamma$ and $\bar\Gamma$ the decay rates of $\nu_R$ into leptons and antileptons respectively,
\be
\begin{aligned}
\Gamma_{a\alpha}&\equiv \Gamma\left(\nu_{Ra}\to\ell_{L\alpha}\phi^*\right)\\
\bar\Gamma_{a\alpha}&\equiv \Gamma\left(\nu_{Ra}\to\bar\ell_{L\alpha}\phi\right)\,,
\end{aligned}
\ee
where $\ell_L$, and $\phi$ stand for the $SU(2)_L$-doublet left-handed (LH) leptons and the $SU(2)_L$-doublet Higgs field, the CP asymmetry parameter is given by 
\be
\varepsilon^{(a)}\equiv\dfrac{\sum_\alpha\Gamma_{a\alpha}-\bar{\Gamma}_{a\alpha}}{\sum_\alpha\Gamma_{a\alpha}+\bar{\Gamma}_{a\alpha}}\,,
\ee
with $\alpha$ ($a$) the flavour (RH neutrino mass) index. The analytic expression of the CP parameter $\varepsilon$ depends on the product $\lambda^\dag \lambda$, with $\lambda$ the Dirac neutrino Yukawa coupling in the mass basis for the RH neutrinos and for the charged leptons. In the convention where the active neutrino mass term is defined by
\be
-\L\supset \dfrac{1}{2}\bar{\nu}_L^c\,m_\nu\,\nu_L+\hc\,,
\ee
where the mass matrix is diagonalised by the PMNS matrix $U$ according to
\be
\hat{m}_\nu=U^T\,m_\nu\,U\,,
\ee
(the ``$\string^$'' symbol is adopted here and in the following to refer to diagonal matrices), $\lambda$ is a matrix in flavour space that can be written in the Casas-Ibarra parametrisation as follows~\cite{Casas:2001sr}:
\be
\lambda=\dfrac{\sqrt{2}}{v}U\hat{m}_\nu^{1/2}R\hat{M}^{1/2}\,,
\label{lambdanuhat}
\ee
where $v=246\GeV$ is the EW vacuum expectation value (VEV), $R$ a complex orthogonal matrix, and $\hat{M}$ the diagonal mass matrix of the RH neutrinos. This expression depends on 9 low-energy parameters, i.e. 3 active neutrino masses, 3 mixing angles, 1 CP violating Dirac phase and 2 CP violating Majorana phases, and on 9 high-energy ones, corresponding to the 3 RH neutrino masses and the 6 parameters of the matrix $R$. The latter is typically independent from the low-energy quantities and its parameters are arbitrary. In general, this prevents to uniquely determine the parameter $\varepsilon$ in terms of low-energy observables and the RH neutrino masses.

The use of flavour symmetries helps improving the predictivity in this scenario: as a flavour symmetry rules the interactions among the different fermion generations, the $R$ matrix might be (partially) fixed, allowing to predict the value of $\varepsilon$ (almost) just in function of neutrino masses, mixings and phases. Some examples can be found in Refs.~\cite{Branco:2001pq,Branco:2002kt,Pascoli:2006ie,Mohapatra:2006se,Pascoli:2006ci,Engelhard:2007kf,
Molinaro:2007uv,Molinaro:2008rg,Jenkins:2008rb,Bertuzzo:2009im,AristizabalSierra:2009ex,Deppisch:2010fr} (see also Refs.~\cite{Hernandez:2015wna,Hernandez:2016kel} for predictive scenario not involving flavour symmetries).

The aim of the present paper is to investigate on a specific scenario where a continuous non-Abelian group is implemented in the Lagrangian as a global flavour symmetry, providing an exceptionally predictive framework for both Leptogenesis and low-energy observables. The symmetry under consideration is the one of the so-called Minimal Flavour Violation (MFV) in the lepton sector (MLFV), considering the type I Seesaw mechanism with three RH neutrinos. The MFV ansatz~\cite{Chivukula:1987py} consists in assuming that any source of flavour and CP violation in any theory Beyond the SM (BSM) is the one in the SM, i.e. the Yukawa couplings. This concept has been technically formulated in terms of the flavour symmetry of the fermion kinetic terms of a given Lagrangian~\cite{DAmbrosio:2002vsn}: the flavour group is a product of a $U(3)$ factor for each fermion in the spectrum, and it is $U(3)^6$~\cite{Cirigliano:2005ck,Cirigliano:2006su,Davidson:2006bd,Gavela:2009cd,Alonso:2011jd} for the type I Seesaw mechanism with 3 RH neutrinos. The Yukawa interactions are the only terms of the renormalisable Lagrangian that are not invariant under the flavour symmetry, unless the Yukawa couplings are promoted to be fields, dubbed spurions, transforming non-trivially under  $U(3)^6$. In the original proposal~\cite{DAmbrosio:2002vsn}, the Yukawa spurions are dimensionless, non-dynamical fields that acquire background values (they could be interpreted as VEVs if the spurions were promoted to be dynamical fields~\cite{Alonso:2011yg,Alonso:2012fy,Alonso:2013mca,Alonso:2013nca}), breaking explicitly the flavour symmetry, and reproducing the measured values of fermion masses and PMNS angles and phases. 

In the quark sector, any non-renormalisable operator containing fermion fields is, eventually, made invariant under the flavour symmetry by the insertion of suitable powers of the Yukawa spurions. Once the latter acquire their background values, the strength of the effects induced by such effective operators is suppressed by specific combinations of quark masses, mixing angles and CP violating phase. In consequence, the cut-off scale that suppresses any non-renormalisable operator can be of the order of a few TeV~\cite{DAmbrosio:2002vsn,Grinstein:2006cg,Hurth:2008jc,Kagan:2009bn,Grinstein:2010ve,Feldmann:2010yp,Guadagnoli:2011id,Buras:2011zb,Buras:2011wi,Alonso:2012jc,Isidori:2012ts,Lopez-Honorez:2013wla,Bishara:2015mha,Lee:2015qra,Arias-Aragon:2017eww}, instead of hundreds of TeV as in the generic case~\cite{Isidori:2010kg}.

In the lepton sector, with the addition of three RH neutrinos, the predictive power of the MLFV is lost in the most generic case. Indeed, three quantities, and not only two as in the quark case, need to be promoted to spurions, i.e. the charged lepton Yukawa, the neutrino Dirac Yukawa and the RH neutrino Majorana mass matrices. A simple parameter counting reveals that it is not possible to uniquely determine the three spurion backgrounds in terms of lepton masses and PMNS parameters. This prevents to link the coefficients of flavour changing operators with low-energy quantities, with the consequent loss of predictivity. A way out is to reduce the symmetry content: in Refs.~\cite{Cirigliano:2005ck,Davidson:2006bd} the non-Abelian part of the $U(3)$ symmetry associated to the RH neutrinos was substituted for a simpler $SO(3)$ plus the hypothesis of CP conservation; in Ref.~\cite{Alonso:2011jd}, instead, it was identified with the one of the lepton $SU(2)_L$ doublets, thus considering a vectorial $SU(3)_V$ flavour symmetry. Both approaches allow to reduce the number of spurions to two, restoring the predictivity of the models: the effects of any flavour changing effective operator can be described in terms of lepton masses and PMNS parameters~\cite{Cirigliano:2005ck,Cirigliano:2006su,Davidson:2006bd,Gavela:2009cd,Alonso:2011jd,Feldmann:2016hvo,Alonso:2016onw}. An updated phenomenological analysis of these two different MLFV realisations has been recently presented in Ref.~\cite{Dinh:2017smk}. A fundamental distinction between them is that the CP conservation hypothesis of the $SO(3)\times\text{CP}$ version is disfavoured by the recent indication of a CP non-conserving Dirac phase in the PNMS matrix~\cite{Forero:2014bxa,Blennow:2014sja,Capozzi:2013csa,Capozzi:2016rtj,Esteban:2016qun,Capozzi:2017ipn,deSalas:2017kay}.

Leptogenesis in the MLFV context has already been investigated in Ref.~\cite{Cirigliano:2006nu} (see also Refs.~\cite{Branco:2006hz,Uhlig:2006xf,Cirigliano:2007hb,Pilaftsis:2015bja}), considering the $SO(3)\times\text{CP}$ version: in order to guarantee a leptonic source of CP violation necessary to explain the measured BAU, the CP conservation hypothesis has been relaxed; in consequence, the precise prediction of flavour effects at low-energy in terms of lepton masses, mixing and phases has been lost. The aim of this paper is to investigate Leptogenesis in the $SU(3)_{V}$ MLFV version introduced in Ref.~\cite{Alonso:2011jd}, where no additional hypothesis on CP  is made and the present indication for the Dirac CP phase of the PMNS can be fulfilled. 

The rest of the paper is structured as follows. The $SU(3)_V$ MLFV scenario under consideration is described in Sect~\ref{Sect:MLFV}. The Leptogenesis CP asymmetry parameter $\varepsilon$ and the Boltzmann equations are discussed in Sect.~\ref{Sect:Lepto}. The numerical results are presented in Sect.~\ref{Sect:NumAna}, showing that a correct value for the BAU is achieved only in a part of the allowed parameter space, testable with (non-)observation of the neutrinoless double beta decay. Concluding remarks can be found in Sect.~\ref{Sect:Conc}.

%
%
\boldmath
\section{The Minimal Lepton Flavour Violation with Vectorial $SU(3)_V$}
\label{Sect:MLFV}
\unboldmath

The use of flavour symmetries to explain the flavour puzzle in the SM goes back to 1978, when Froggatt and Nielsen~\cite{Froggatt:1978nt} first introduced a single $U(1)$ factor to describe the quark mass hierarchies and the CKM mixing matrix. Subsequent analyses also included the lepton sector~\cite{Altarelli:2000fu,Altarelli:2002sg,Chankowski:2005qp,Buchmuller:2011tm,Altarelli:2012ia,Bergstrom:2014owa}, where however a larger freedom is present due to the lack of knowledge of some neutrino parameters. At the beginning of this century, the use of flavour discrete symmetries became very popular due to the high predictive power in the lepton sector of this kind of models~\cite{Ma:2001dn,Babu:2002dz,Altarelli:2005yp,Altarelli:2005yx}. These constructions have been later extended to the quark sector, attempting to provide a unified explanation of the flavour puzzle~\cite{Feruglio:2007uu,Bazzocchi:2009pv,Bazzocchi:2009da,Altarelli:2009gn,Toorop:2010yh,Altarelli:2010gt,Toorop:2010yh,Varzielas:2010mp,Toorop:2011jn,Grimus:2011fk,deAdelhartToorop:2011re,King:2011ab,Altarelli:2012ss,Bazzocchi:2012st,King:2013eh}, and they have been shown to be contexts where flavour violating processes are under control with new physics at the TeV scale~\cite{Feruglio:2008ht,Feruglio:2009iu,Lin:2009sq,Feruglio:2009hu,Ishimori:2010au,Toorop:2010ex,Toorop:2010kt,Merlo:2011hw,Altarelli:2012bn}. Only in 2011, with the discovery of a non-vanishing and relatively large leptonic reactor angle~\cite{Abe:2011sj,Adamson:2011qu,Abe:2011fz,An:2012eh,Ahn:2012nd}, strong doubts raised on the goodness of non-Abelian discrete models to describe Nature.

In this panorama, the idea of MFV\footnote{Despite being so predictive, the MFV only describes masses and mixings, but does not explain their origin: indeed, no justification is provided for the Yukawa spurion background. Improvements with this respect can be found in Refs.~\cite{Alonso:2011yg,Alonso:2012fy,Alonso:2013mca,Alonso:2013nca} (see also Refs.~\cite{Anselm:1996jm,Barbieri:1999km,Berezhiani:2001mh,Feldmann:2009dc,Nardi:2011st}).} experienced a new revival of interest: this context is more predictive than models based on the Froggatt-Nielsen $U(1)$ and escapes from the rigidity of the discrete constructions. This section will summarise the main aspects of the MLFV scenario presented in Ref.~\cite{Alonso:2011jd}, fixing at the same time the notation used throughout this paper.

Considering the SM spectrum supplemented with three RH neutrinos, the flavour symmetry characterising the $SU(3)_V$ MLFV scenario is $\G_F^\text{NA}\times \G_F^\text{A}$ where
\be
\begin{aligned}
\G_F^\text{NA}&\equiv SU(3)_{V}\times SU(3)_{e_R}\\
\G_F^\text{A}&\equiv U(1)_Y\times U(1)_L \times U(1)_{e_R}\,.
\end{aligned}
\label{FlavourSymmetries}
\ee
The distinction between Abelian and non-Abelian terms reflects the fact that the non-Abelian symmetry factors deal exclusively with the inter-generation hierarchies~\cite{Alonso:2011yg,Alonso:2012fy,Alonso:2013mca,Alonso:2013nca}, while the Abelian ones may explain the hierarchies between the third generation fermions, such as the ratio $m_\tau/m_t$. 
The choice of $\G_F^\text{A}$ in Eq.~(\ref{FlavourSymmetries})  is the result of using the freedom of rearranging the $U(1)$ factors in order to identify the hypercharge, the Lepton number and transformations that act globally on the RH charged lepton fields only. 

The part of the Lagrangian containing the kinetic terms is invariant under $\G_F^\text{NA}\times \G_F^\text{A}$ with the lepton field transformations under $\G_F^\text{NA}\times U(1)_L \times U(1)_{e_R}$ as
\be
\ell_L\sim ({\bf3},\,1)_{(1,0)}\qquad
e_R\sim(1,\,{\bf3})_{(1,1)}\qquad
\nu_R\sim({\bf3},\,1)_{(1,0)}\,,
\label{TransformationsLeptons}
\ee
where $e_R$ are the RH charged leptons. Instead, this is not the case for the part describing the lepton masses. The Type I Seesaw Lagrangian, which can be written as~\cite{Alonso:2011jd}
\be
-\L=\epsilon_e \ov{\ell}_L\phi Y_e e_R+\ov{\ell}_L\widetilde\phi Y_\nu \nu_R
+\dfrac12\mu_L \bar{\nu}_R^c Y_M \nu_R+\hc\,,
\label{YukawaLagrangian}
\ee
describes the light active neutrino masses at low-energy through the so-called Weinberg operator~\cite{Weinberg:1979sa},
\be
\O_5=\dfrac{1}{2}\left(\bar{\ell}_L\tilde\phi\right)Y_\nu\dfrac{Y_M^{-1}}{\mu_L}Y_\nu^T\left(\tilde\phi^T\ell^c_L\right)+\hc\,,
\ee
where $\widetilde\phi\equiv i\sigma_2\phi^*$, $Y_e$, $Y_\nu$ and $Y_M$ are $3\times3$ matrices in the flavour space, $\mu_L$ is the scale of lepton number violation and $\epsilon_e$ is a constant that will be associated to the breaking of the $U(1)_{e_R}$ symmetry. By the first Shur's lemma, as $\ell_L$ and $\nu_R$ transform as triplets under the same symmetry factor, $Y_\nu$ is necessarily a unitary matrix and can be redefined away with a $SU(3)_V$ transformation~\cite{Bertuzzo:2009im,Alonso:2011jd}: this Dirac Yukawa term is then invariant under $\G_F^\text{NA}\times\G_F^\text{A}$. On the contrary, the charged lepton Yukawa interactions break $\G_F^\text{NA}\times U(1)_{e_R}$ and the RH neutrino masses break $SU(3)_V\times U(1)_L$. A way-out to recover the invariance of the whole mass Lagrangian is to promote the two Yukawa matrices $Y_e$ and $Y_M$, and the two parameters $\epsilon_e$ and $\mu_L$ to be spurion fields, i.e. non-dynamical fields that transform non-trivially under $\G_F^\text{NA}\times \G_F^\text{A}$. Selecting the spurion transformations under $\G_F^\text{NA}$ as
\be
Y_e\sim({\bf3},\,{\bf\ov3})\qquad\qquad
Y_M\sim({\bf\ov6},\,1)
\label{NonAbelianSpurions}
\ee
and under $U(1)_L \times U(1)_{e_R}$ as
\be
\mu_L\sim(-2,\,0)
\qquad\qquad
\epsilon_e\sim(0,\,-1)\,,
\label{AbelianSpurions}
\ee
the mass Lagrangian is formally invariant under the entire flavour symmetry.

Lepton masses and mixings arise only once the spurion fields acquire background values, breaking explicitly the flavour symmetry: in the charged lepton mass basis,
\be
\begin{aligned}
\hat{Y}_e&=\dfrac{\sqrt2}{\epsilon_e v}\diag(m_e,\,m_\mu,\,m_\tau)\\
Y_M&=\dfrac{v^2}{2\mu_L}U^* \hat{m}_\nu^{-1}U^\dag\,,
\end{aligned}
\label{NonAbelianSpurionsBG}
\ee
where $\epsilon_e$ and $\mu_L$ are respectively a dimensionless quantity and a mass. Notice that the same symbols have been used for the couplings in Eq.~(\ref{YukawaLagrangian}), for the spurions in Eqs.~(\ref{NonAbelianSpurions}) and (\ref{AbelianSpurions}), and for their background values in Eq.~(\ref{NonAbelianSpurionsBG}): it will be clear which is the meaning associated to each symbol in the formulae that follow.

An estimate of $\epsilon_e$ and of $\mu_L$ follows by assuming that the largest eigenvalues of $Y_e$ and of $Y_M$ are $\lesssim1$~\footnote{Considering values larger than 1 would imply that multiple products of Yukawa spurions would be more relevant than the single spurions themselves, and then they should be treated in a non-perturbative approach~\cite{Kagan:2009bn}.}: then
\be
\begin{aligned}
\epsilon_e\sim&\dfrac{\sqrt2 m_\tau}{v}\approx0.01\\
\mu_L\gtrsim&\dfrac{v^2}{2\sqrt{\Delta m^2_\text{atm}}}\approx6\times 10^{14}\GeV\,,
\end{aligned}
\label{BoundsE_eMu_L}
\ee
where $\Delta m^2_\text{atm}\approx 2.5\times 10^{-3}\eV^2$~\cite{Esteban:2016qun,Capozzi:2017ipn} is the atmospheric squared mass difference of the light active neutrinos and the ``$\gtrsim$'' symbol reflects the fact that the absolute neutrino mass scale is still unknown. Within this setup, the expected mass scale of the RH neutrinos is of order $\mu_L$.

In the spirit of the MLFV, any non-renormalisable operator can be made invariant under the flavour symmetry by inserting suitable combinations of the spurions. Once the latter acquire background values, the strength of each operator gets suppressed by a combination of lepton masses and PMNS parameters. These extra suppressions allow to predict the rates for rare radiative lepton decays and lepton conversion in nuclei in agreement with present data with a new physics scale that suppresses the effective operators as low as the TeV (see Ref.~\cite{Dinh:2017smk} for a recent update).

Spurion insertions can be introduced not only in effective operators, but also in the renormalisable terms of the Lagrangian~\footnote{Some operators that are non-renormalisable in the description considered here appear in the list of the renormalisable ones if a non-SM Higgs field is considered, as described in the so-called Higgs Effective Field Theory~\cite{Feruglio:1992wf,Grinstein:2007iv,Contino:2010mh,Alonso:2012px,Alonso:2012pz,Buchalla:2013rka,Brivio:2013pma,Gavela:2014vra,Alonso:2014wta,Hierro:2015nna,Gavela:2014uta,Gavela:2016bzc,Eboli:2016kko,Brivio:2016fzo,deFlorian:2016spz,Merlo:2016prs}. As shown in Ref.~\cite{Alonso:2012pz,Brivio:2013pma,Brivio:2014pfa,Gavela:2014vra,Brivio:2015kia,Brivio:2016fzo,Brivio:2017ije,Hernandez-Leon:2017kea,Merlo:2017sun}, a different phenomenology is expected with a non-SM Higgs in the spectrum. In the present paper, however, the standard formulation with a $SU(2)_L$-doublet Higgs is retained.}. In particular, the introduction of spurions in the Dirac Yukawa term will be shown to be necessary in order to achieve  successful Leptogenesis. Considering only the most relevant contributions, the Dirac Yukawa term can be written as
\be
\ov{\ell}_L\widetilde\phi \left(\unity+c_1 \hat{Y}_e \hat{Y}_e^\dag +c_2 Y_M^\dag Y_M\right) \nu_R\,,
\ee
where $c_{1,2}$ are dimensionless real parameters that are taken to be smaller than 1 in order to enforce a perturbative approach\footnote{In Ref.~\cite{Branco:2006hz}, considering the $SO(3)\times CP$ version of MLFV, the equivalent of the coefficients $c_{1,2}$ have been shown to be generated by radiative corrections during the evolution of the Lagrangian parameters.}. Within this hypothesis, the expression for $Y_M$ in Eq.~(\ref{NonAbelianSpurionsBG}) holds in first approximation.

\subsection{A Suitable Basis for Leptogenesis}

The explicit computation of the $\varepsilon$ parameter that controls the amount of CP asymmetry generated in the RH neutrino decays is typically performed in the mass basis for charged leptons and for RH neutrinos. The mass Lagrangian in this basis reads
\be
-\L=\epsilon_e \ov{\ell}_L\phi \hat{Y}_e e_R+\ov{\ell}_L\widetilde\phi\, \lambda\, \nu_R
+\dfrac12\mu_L \bar{\nu}_R^c \hat{Y}_M \nu_R+\hc\,,
\label{YukawaLagrangianRedef}
\ee
where $\lambda$ is the Dirac neutrino Yukawa in this basis. Considering the background values of the spurions in Eq.~(\ref{NonAbelianSpurionsBG}), $\lambda$ reads
\be
\lambda=U\left(\unity+c_1 U^\dag\hat{Y}^2_e U + c_2 \hat{Y}_M^2\right)\,,
\label{lambda}
\ee
where $\hat{Y}_e$ is defined in Eq.~(\ref{NonAbelianSpurionsBG}), while
\be
\hat{Y}_M=\dfrac{v^2\hat{m}^{-1}_\nu}{2\mu_L}\,.
\label{BackgroundYM}
\ee
The two parameters $c_1$ and $c_2$ control the complex contributions coming from the PMNS matrix and the real contributions coming from the diagonal RH neutrino mass matrix, respectively. They are expected to be of the same order of magnitude and they will be taken equal to each other in what follows in order to simplify the study of the parameter space. It will be shown {\it a posteriori} that relaxing this condition has not relevant impact on the results as far as they are taken of the same order of magnitude.

The relevance of the spurion insertions becomes evident computing the value of three specific weak-base invariants~\cite{Branco:2001pq}, related to the CP violation responsible for Leptogenesis:
\be
\begin{aligned}
\I_1=&\Im\left(\tr\left[\lambda^\dag\lambda \hat{Y}_M^3\lambda^T\lambda^*\hat{Y}_M\right]\right)\\
\I_2=&\Im\left(\tr\left[\lambda^\dag\lambda \hat{Y}_M^5\lambda^T\lambda^*\hat{Y}_M\right]\right)\\
\I_3=&\Im\left(\tr\left[\lambda^\dag\lambda \hat{Y}_M^5\lambda^T\lambda^*\hat{Y}_M^3\right]\right)\,.\\
\end{aligned}
\ee
It is straightforward to show that the three invariants depend on the combinations $\Im\left[\left(\lambda^\dag\lambda\lambda^\dag\lambda\right)_{\alpha\neq \beta}\right]$: if $\lambda$ was taken without the spurions insertions, then $\lambda=U$ and the three invariants together with the parameter $\varepsilon$ would vanish.

%
\section{Baryogenesis trough Leptogenesis}
\label{Sect:Lepto}

The prediction for the baryon asymmetry in the Universe requires to compute the CP asymmetry parameter $\varepsilon$ and to take into consideration its evolution during the expansion of the Universe, which depends on the interactions that are in thermal equilibrium at different temperatures. With this respect, the value of the RH neutrino mass scale $\mu_L$ is a fundamental parameter as it identifies different flavour regimes~\cite{Barbieri:1999ma,Pilaftsis:2005rv,Abada:2006fw,Nardi:2006fx,Underwood:2006xs,Abada:2006ea,Blanchet:2006be}: the lower $\mu_L$ is, the more relevant the flavour composition of the charged leptons produced in the RH neutrino decays is. For the $SU(3)_V$ MLFV framework, $\mu_L\gtrsim10^{14}\GeV$ and it corresponds to the so-called unflavoured regime, where the charged lepton flavour does not play any role. Indeed, the only relevant interactions at these energies are the Yukawa ones, which induce RH neutrino decays, and the gauge ones that are flavour blind: lepton and anti-lepton quantum states propagate coherently between the production in decays and the later absorption from inverse decays. 

In addition, the scale $\mu_L$ identifies the reheating temperature necessary for the thermal production of the RH neutrinos~\cite{Giudice:2003jh,Buchmuller:2004nz}: once the temperature drops below $M_a$, the thermal production of the corresponding RH neutrino $N_a$ becomes irrelevant. This allows to identify a lower bound on the reheating temperature at about $10^{13\div14}\GeV$ in the MLFV scenario under consideration. The usually quoted upper bound of $10^{6\div10}\GeV$ does not apply as it is exclusively connected to the so-called gravitino problem in supersymmetry~\cite{Khlopov:1984pf,Ellis:1984eq,Kohri:2005wn}.

Besides $\mu_L$, the splitting between the RH neutrino masses is also relevant: when the spectrum is highly hierarchical then the asymmetries produced by the heaviest states are typically (partially) washed out by the inverse decay of the lightest states (i.e. $\ell_{L\alpha}(\bar\ell_{L\alpha})+\phi^\ast (\phi)\to \nu_{Ra}$) and by the $2\leftrightarrow2$ scattering mediated by the lightest states (i.e. $\ell_{L\alpha}+\phi^\ast\leftrightarrow\bar{\ell}_{L\alpha}+\phi$); when instead the spectrum is degenerate, a resonance in the decay rate is present~\cite{Liu:1993tg,Flanz:1994yx,Flanz:1996fb,Covi:1996wh,Covi:1996fm,Pilaftsis:1997jf,Buchmuller:1997yu,Pilaftsis:2003gt}, which, however, is diluted due to the washout effects of all the three RH neutrinos. In the framework under consideration, depending on the mass of the lightest active neutrino, the spectrum varies from hierarchical to degenerate and therefore the computation of $\eta_B$ is not straightforward. In particular, when the heavier RH neutrinos also contribute to the final asymmetry, the flavour composition of the three RH neutrinos is relevant and need to be taken into consideration~\cite{Barbieri:1999ma,Engelhard:2006yg,Bertuzzo:2009im,Antusch:2010ms,Blanchet:2011xq}: part of the asymmetry generated by a heavier RH neutrino may escape the washout from a lighter one; moreover, part of the final asymmetry may not come from the production in the RH neutrino decays, but from the dilution effects. The density matrix formalism~\cite{Barbieri:1999ma,Abada:2006fw,Blanchet:2006ch,DeSimone:2006nrs,Blanchet:2011xq,Blanchet:2012bk} (see Ref.~\cite{Dev:2014laa} for an alternative flavour-covariant formalism) turns out to be extremely effective in these cases, and thus for the MLFV framework under discussion: it allows to calculate the asymmetry not only in the well definite regimes with a hierarchical or degenerate RH neutrino spectrum, but also in the intermediate cases, describing together the lepton quantum states and the thermal bath. 

In the rest of this section, the density matrix approach will be adopted following Ref.~\cite{Blanchet:2011xq}, fixing the notation and illustrating the procedure to follow, while in the next section the results of the numerical simulation will be presented. In the present analysis several contributions will not be considered, as their impact is not relevant for the results presented here: they are due to $\Delta L=1$ scatterings~\cite{Luty:1992un,Plumacher:1996kc,Buchmuller:2000as,HahnWoernle:2009qn}, thermal corrections~\cite{Kiessig:2010pr,Giudice:2003jh}, momentum dependence~\cite{Basboll:2006yx,HahnWoernle:2009qn}, and quantum kinetic effects~\cite{DeSimone:2007gkc,Beneke:2010wd,Anisimov:2010dk,Dev:2014wsa}.\\

The baryon-to-photon number ratio at recombination, whose best experimental determination is reported in Eq.~(\ref{ExpBAU}), can be written in terms of the final $B-L$ asymmetry density $N_{B-L}^f$ as
\be
\eta_B=c_s\dfrac{N_{B-L}^f}{N_\gamma^\text{rec}}\simeq0.0096\, N_{B-L}^f\,,
\ee
with $c_s=28/79$ defined in Eq.~(\ref{cs}) for $N_F=3$, and $N_\gamma^\text{rec}\simeq37$ the photon number density at recombination. 

The final $B-L$ asymmetry results from the sum of the asymmetries generated by the three RH neutrinos, in case partially washed out by the inverse decays~\cite{Barbieri:1999ma,Strumia:2006qk}. It can be calculated solving the following system of four differential equations:
\be
\hspace{-5mm}
\begin{aligned}
\frac{\d (N_{B-L})_{\alpha \beta}}{\d z} =& \varepsilon_{\alpha \beta}^{(a)}\D_a[z]\left(N_{N_a}-N_{N_a}^\text{eq}\right)-\frac{\W_a[z]}{2}\left\lbrace \P^{(a)0},\, N_{B-L}\right\rbrace _{\alpha \beta}\\
\frac{\d N_{N_a}}{\d z} =& -\D_a[z]\left(N_{N_a}-N_{N_a}^\text{eq}\right)\qquad (a=1,2,3)\,.
\end{aligned}
\label{BEs}
\ee
The parameter $z$ is the ratio between the lightest RH neutrino mass $\Ml$ and the temperature of the bath, i.e. $z\equiv \Ml/T$. $N_X$ is any particle number or asymmetry $X$ calculated in a portion of co-moving volume containing one RH neutrino in ultra-relativistic thermal equilibrium, that is $N_{N_a}^\text{eq}[z\ll1]=1$. The expression for $N_{N_a}^\text{eq}[z_a]$ at a $z_a\equiv \sqrt{x_a}z$, with $x_a=M_a^2/M^2_\text{light}$, is given by~\cite{Buchmuller:2004nz,Blanchet:2011xq}
\be
N_{N_a}^\text{eq}[z_a]=\dfrac12 z_a^2\,\K_2[z_a]\,,
\ee
where $\K_n[z_a]$ is a modified Bessel function, satisfying to
\be
z_a^2\, y^{\prime\prime}+z_a\,y^\prime-(z_a^2+n^2)y=0\,.
\ee
The $\D_a[z]$ terms are the RH neutrino decay factors~\cite{Kolb:1990vq}
\be
\D_a[z]\equiv\dfrac{\Gamma^\D_a[z_a]}{H[z_a]\, z}=K_a \, x_a\,z\,\dfrac{\K_1[z_a]}{\K_2[z_a]}\,,
\label{DecayFactors}
\ee
where the total decay rates $\Gamma^\D_a[z_a]$ read~\cite{Kolb:1979qa}
\be
\Gamma^\D_a[z_a]\equiv \left(\sum_\alpha\Gamma_{a\alpha}+\bar{\Gamma}_{a\alpha}\right)\dfrac{\K_1[z_a]}{\K_2[z_a]}\,,
\ee
where $\K_1[z_a]$ is also a modified Bessel function, and $H[z_a]$ is the Hubble expansion rate of the Universe given by
\be
H[z_a]\equiv\sqrt{\dfrac{8\pi^3g_\star}{90}}\dfrac{M_a^2}{M_\text{Pl}}\dfrac{1}{z_a^2}\,,
\ee
where $g_\star=g_\text{SM}=106.75$ is the total number of degrees of freedom and $M_{Pl}=1.22\times 10^{19}\GeV$ the Planck mass. The second expression on the right-hand side of Eq.~(\ref{DecayFactors}) contains the total decay parameters $K_a$ that measure the strength of the washout: they are defined  as the ratio between the total decay widths of the RH neutrinos calculated at a temperature much smaller than $M_a$ and the Hubble parameter at $T=M_a$, when the RH neutrinos start to become non-relativistic: explicitly,
\be
K_a\equiv \dfrac{\left(\Gamma_a+\bar{\Gamma}_a\right)_{z_a\gg1}}{H[z_a=1]}\,,
\ee
where
\be
\Gamma_a\equiv \sum_\alpha \Gamma_{a\alpha}\qquad\qquad
\bar\Gamma_a\equiv \sum_\alpha \bar\Gamma_{a\alpha}\,.
\ee
For $K_a\gg1$ the RH neutrinos decay and inverse-decay many times before entering the non-relativistic regime: in consequence their abundance is close to the equilibrium distribution and this case is dubbed strong washout regime. On the other side, for $K_a\ll1$, called weak washout regime, the majority of the RH neutrinos decay completely out-of-equilibrium, already in the non-relativist regime, and therefore their equilibrium abundance is exponentially suppressed by the Boltzmann factor. Introducing the notation of the so-called effective washout parameter~\cite{Plumacher:1996kc} and of the equilibrium neutrino mass~\cite{Nezri:2000pb,Buchmuller:2003gz,Buchmuller:2004tu},
\be
\begin{aligned}
\widetilde{m}_a&\equiv \dfrac{v^2}{2}\dfrac{(\lambda^\dag \lambda)_{aa}}{M_a}\,,\\
m_\star&=\dfrac{16\pi^{5/2}\sqrt{g_\star}}{3\sqrt5}\dfrac{v^2}{2M_{Pl}}\simeq1.07\times 10^{-3}\eV\,,
\end{aligned}
\ee
the total decay parameter can be written as 
\be
K_a=\dfrac{\widetilde{m}_a}{m_\star}\,.
\ee

The $\W_a[z]$ terms are the washout factors due to inverse decays~\cite{Kolb:1979qa,Dolgov:1981hv,Giudice:2003jh}
and $\Delta L=2$ processes~\cite{Dolgov:1981hv,Kolb:1979qa,Giudice:2003jh}, which provide the two relevant effects for these values of the RH neutrino masses:
\be
\W_a[z]\equiv \W^{ID}_a[z] + \Delta \W_a[z]\,,
\ee
where the two factors are defined as
\be
\begin{aligned}
\W^{ID}_a[z]&=\dfrac14\, K_a\,\K_1[z_a]\,z_a^3\\
\Delta \W_a[z]&\simeq \dfrac{\alpha}{z_a^2}\,M_a\,\widetilde{m}_a^2\qquad\text{for}\qquad z_a\gg1\,,
\end{aligned}
\ee
with $\alpha$ given by~\cite{Bertuzzo:2009im}
\be
\alpha=\dfrac{3\sqrt5 M_{Pl}}{\zeta(3)\pi^{9/2}v^4\sqrt{g_\star}}\,,
\label{alphaexpression}
\ee
being $\zeta(3)\approx1.202$ the Ap\'ery constant. The inverse decay processes are relevant when they are in equilibrium, i.e. $\W_a[z]\gtrsim1$, and this occurs only in the strong washout regime for $K_a>3$. Instead, in the weak washout regime, $\W_a[z]<1$ and the inverse decays are always irrelevant. On the other side, the $\Delta L=2$ processes have a relevant effect only for $z\gtrsim z_\Delta\gg1$, where $z_\Delta$ is determined by
\be
\W^{ID}_a[z_\Delta]=\Delta \W_a[z_\Delta]\,.
\label{zDelta}
\ee

The $\P^{(a)0}$ factors are the flavour projectors along the $\ell_a$ direction defined by
\be
\P^{(a)0}_{\alpha\beta}=\dfrac{\lambda^\ast_{\beta a}\lambda_{\alpha a}}{\left(\lambda^\dag\lambda\right)_{aa}}\,,
\ee
where $\lambda$ is the Dirac Yukawa in Eq.~(\ref{lambda}). The suffix ``$0$'' indicates that only the leading terms are considered.

Finally, the flavoured CP asymmetry parameters $\varepsilon_{\alpha \beta}^{(a)}$ are given by~\cite{Covi:1996wh,Liu:1993tg,Flanz:1994yx,Flanz:1996fb,Covi:1996wh,Covi:1996fm,Pilaftsis:1997jf,Buchmuller:1997yu,Pilaftsis:2003gt}

\begin{widetext}
\begin{equation}
\hspace{-2cm}
\begin{split}
\epsilon_{\alpha \beta}^{(a)} =  \frac{i}{16\pi (\lambda^{\dagger}\lambda)_{aa}} \sum_{b\neq a}&
\left\{\left[\lambda_{\alpha a}\lambda_{\beta b}^*(\lambda^{\dagger}\lambda)_{ba} 
- \lambda^*_{\beta a}\lambda_{\alpha b} (\lambda^{\dagger}\lambda)_{ab}\right]
\frac{M_b}{M_a} \left[\left(1+\frac{M_b^2}{M_a^2}\right)\ln \left(1+\frac{M_a^2}{M_b^2}\right)-\frac{M_a^2(M_a^2-M_b^2)}{(M_a^2-M_b^2)^2+\left(M_a\Gamma_a+M_b\Gamma_b\right)^2}-1\right]+\right.\\
&\hspace{5mm} \left.-\left[\lambda_{\alpha a}\lambda_{\beta b}^*(\lambda^{\dagger}\lambda)_{ab}-\lambda^*_{\beta a}\lambda_{\alpha b}(\lambda^{\dagger}\lambda)_{ba}\right]\frac{M_a^2(M_a^2-M_b^2)}{(M_a^2-M_b^2)^2+\left(M_a\Gamma_a+M_b\Gamma_b\right)^2}\right\}\,,
\end{split}
\label{newepsilon}
\end{equation}
\end{widetext}
where the Kadanoff-Bayn regulator~\cite{Garny:2011hg}, that is the term in the denominator containing the RH neutrino decay rates $\Gamma_a$, plays an important role when the spectrum is almost degenerate. Different regulators can be considered, depending on the formalism chosen: the one used in the previous expression prevents the arising of any divergence in the doubly degenerate limit $M_a\to M_b$ and $\Gamma_a\to \Gamma_b$, which instead occurs within the classical Boltzmann approach.

%
%
\section{Numerical Analysis}
\label{Sect:NumAna}

This section contains the results of the numerical analysis first focussing on the baryon asymmetry and then on the neutrinoless double beta decay. 

The input data used are the PDG values for the charged lepton masses~\cite{Olive:2016xmw}
\be
\begin{aligned}
m_e=&0.51\MeV\,,\\
m_\mu=&105.66\MeV\,,\\
m_\tau=&1776.86\pm0.12\MeV\,,
\end{aligned}
\ee
where the electron and muon masses are shown without errors as the sensitivities are completely negligible, and the results of the neutrino oscillation fit from Ref.~\cite{Esteban:2016qun} (see also Ref.~\cite{Capozzi:2017ipn,deSalas:2017kay}) reported in Table~\ref{TableOscFit}. In the analysis that follows, all these input parameters have been taken at their central values.

\begin{table}[htp]
\begin{center}
\begin{tabular}{l|c|c|}
										& Normal Ordering 				& Inverted Ordering \\[1mm]
\hline
&&\\
$\sin^2\theta_{12}$						& $0.307^{+0.013}_{-0.012}$				& $0.307^{+0.013}_{-0.012}$ \\[3mm]
$\sin^2\theta_{23}$						& $0.565^{+0.025}_{-0.120}$				& $0.572^{+0.021}_{-0.028}$ \\[3mm]
$\sin^2\theta_{13}$						& $0.02195^{+0.00075}_{-0.00074}$			& $0.02212^{+0.00074}_{-0.00073}$\\[3mm]
$\delta^\ell_{CP}/^\circ$					& $228^{+51}_{-33}$						& $281^{+30}_{-33}$\\[3mm]
$\Delta m^2_{sol}/10^{-5}\eV^2$			& $7.40^{+0.21}_{-0.20}$					& $7.40^{+0.21}_{-0.20}$\\[3mm]
$\Delta m^2_{atm}/10^{-3}\eV^2$			& $2.515\pm0.035$						& $2.483^{+0.034}_{-0.035}$\\[3mm]
\hline
\end{tabular}
\end{center}
\caption{\footnotesize\it Three-flavour oscillation parameters from the global fit in Ref.~\cite{Esteban:2016qun}. The second and third columns refer to the NO and IO, respectively. The notation chosen is $\Dmsol\equiv m^2_{\nu_2}- m^2_{\nu_1}$ and $\Dmatm\equiv m^2_{\nu_3}-m^2_{\nu_1}$ for NO and $\Dmatm\equiv m^2_{\nu_2}-m^2_{\nu_3}$ for IO. The errors reported correspond to the $1\sigma$ uncertainties.}
\label{TableOscFit}
\end{table}%

Tab.~\ref{TableOscFit} reports the value of the mixing angles and of the Dirac CP phase according to the PDG parametrisation of the PMNS matrix,
\be
U=\sR_{23}(\theta_{23})\,\sR_{13}(\theta_{13},\,\delta^\ell_{CP})\,\sR_{12}(\theta_{12})\,\sP\,,
\ee
where $\sR_{ij}$ is a $3\times 3$ rotation in the flavour space in the $ij$ sector of an angle $\theta_{ij}$ and $\sP$ is the diagonal matrix containing the Majorana CP phases defined by
\be
\sP=\diag\left(1,\,e^{i\frac{\alpha_{21}}{2}},\,e^{i\frac{\alpha_{31}}{2}}\right)\,.
\ee
Tab.~\ref{TableOscFit} also contains the neutrino mass square differences, while the value of the lightest neutrino mass is presently unknown. Moreover, it is still an open issue the ordering of the neutrino mass eigenstates: the so-called Normal Ordering (NO) refers to the case when $m_{\nu_1}<m_{\nu_2}\ll m_{\nu_3}$ while the Inverse Ordering (IO) to the case when $m_{\nu_3}\ll m_{\nu_1}<m_{\nu_2}$. The labelling of the three $\nu_i$ is determined by the flavour content of each mass eigenstate: $\nu_1$ is the state with the largest contamination of $\nu_e$; $\nu_2$ is the one with an almost equally composition of the three flavours; $\nu_3$ is the one almost exclusively defined as a equal mixture of $\nu_\mu$ and $\nu_\tau$. The diagonal active neutrino mass matrix can thus be written in terms of the lightest neutrino mass as follows: for the NO and IO respectively,
\be
\hspace{-3mm}
\begin{aligned}
\hat{m}^\text{NO}_\nu&=
\left(\begin{array}{ccc}
   m_{\nu_1} 	& 0 						& 0 \\
   0 			& \sqrt{m^2_{\nu_1}+\Dmsol} 	& 0 \\
   0 			& 0						& \sqrt{m^2_{\nu_1}+\Dmatm} \\
\end{array}\right)\\
\hat{m}^\text{IO}_\nu&=
\left(\begin{array}{ccc}
   \sqrt{m^2_{\nu_3}+\Dmatm-\Dmsol} 	& 0 						& 0 \\
   0 								& \sqrt{m^3_{\nu_3}+\Dmatm} 	& 0 \\
   0 								& 0						& m_{\nu_3} \\
\end{array}\right)\,.
\end{aligned}
\ee

To match with the notation typically adopted in Leptogenesis, a different convention is chosen for the labelling of the RH neutrino mass eigenstates. For both NO and IO, $N_1$ always refers to the lightest eigenstate, $N_2$ to the next to lightest and the $N_3$ to the heaviest. In consequence, $\hat Y_M$ in Eq.~(\ref{BackgroundYM}) takes a different definition in terms of the three RH neutrino masses depending on the ordering of the spectrum: for the NO and IO respectively,
\be
\begin{aligned}
\mu_L\hat{Y}^\text{NO}_M&\equiv\diag(M_3,\,M_2,\,M_1)\\
\mu_L\hat{Y}^\text{IO}_M&\equiv\diag(M_2,\,M_1,\,M_3)\,.
\end{aligned}
\label{RHNeutrinosMassOrdering}
\ee

\begin{figure}[h!]
\includegraphics[width=0.45\textwidth]{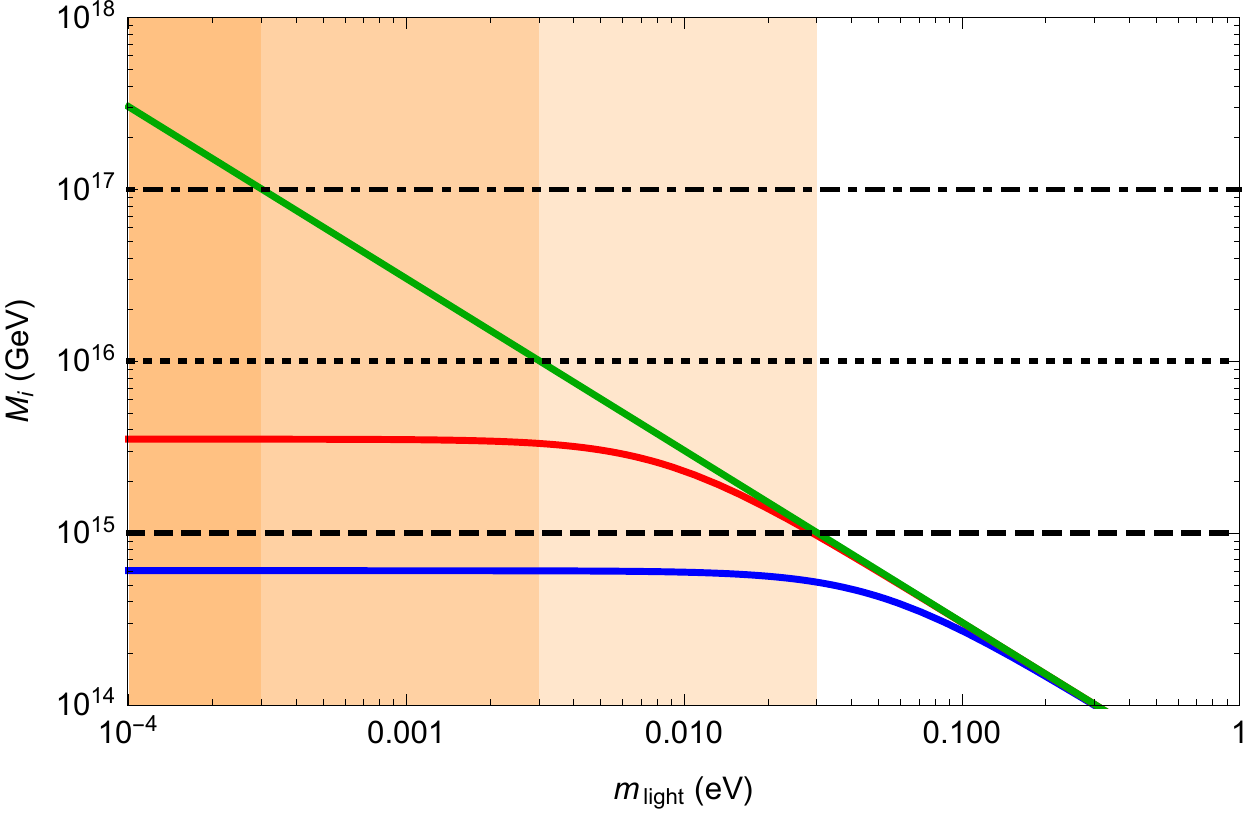}
\caption{\footnotesize\it Profiles of the RH neutrino masses as a function of the lightest active neutrino mass $\ml$. The blue (red) [green] continuous line corresponds to the lightest (next-to-lightest) [heaviest] RH neutrino. The horizontal lines represent different values for the lepton number violation scale: the dashed one refers to $\mu=10^{15}\GeV$, while the dotted to $\mu=10^{16}\GeV$, and the dot-dashed to $\mu=10^{17}\GeV$. The shaded areas are regions where the condition $(\hat Y_M)_{ii}\leq1$ does not hold: three specific cases are illustrated for $\mu_L=10^{15}\GeV$, $10^{16}\GeV$, $10^{17}\GeV$.}
\label{Fig.RH-ActiveMasses}
\end{figure}

The lepton number violation scale $\mu_L$, the spurion background value $\hat Y_M$ and the active neutrino masses are linked together by Eq.~(\ref{BackgroundYM}). In consequence it is possible to identify a range of values for the lightest neutrino mass, given a value for the scale $\mu_L$ and requiring that the largest entry of $\hat Y_M$ is of order 1, according to the MLFV construction illustrated in Sect.~\ref{Sect:MLFV}. Fig.~\ref{Fig.RH-ActiveMasses} shows the profiles of the RH neutrino masses as a function of the lightest active neutrino mass $\ml$ for a NO spectrum. The plot for the IO case is very similar: the only difference is that the line corresponding to the next-to-lightest RH neutrino (in red) almost overlaps with the one of the lightest (in blue). The horizontal lines represent different values for the $\mu_L$ scale, $\mu_L=10^{15},\,10^{16},\,10^{17}\GeV$, and their crossing with the line of the heaviest RH neutrino mass (in green) identifies the lowest value that $\ml$ can take satisfying $(\hat Y_M)_{ii}\leq1$.

Fig.~\ref{Fig.RH-ActiveMasses} shows that the lower bound on $\mu_L$ reported in Eq.~(\ref{BoundsE_eMu_L}) corresponds to the lightest RH neutrino line (in blue) for $\ml\lesssim0.03\eV$. An upper bound on $\mu_L$ can be taken, in full generality, to be at the Planck scale. However, such a large $\mu_L$ is not consistent with the hypothesis of thermal production of RH neutrinos, as the temperature of the Universe should be at least of the same order of magnitude as their masses. In the numerical analysis that follows, the lepton number violation scale is taken at $\mu_L=10^{16}\GeV$: the corresponding heaviest RH neutrino mass satisfies $M_3<10^{16}\GeV$ and the range of values for the lightest active neutrino mass is $\ml\in[0.003,\,0.2]\eV$. In consequence, as shown in Fig.~\ref{Fig.RH-ActiveMasses}, all the three RH neutrinos may contribute to the baryon asymmetry. Further discussion on the maximal temperature of the Universe and on the thermal production of the RH neutrinos will follow at the end of next section.

\subsection{Baryon asymmetry in the Universe}

This subsection is devoted to illustrate the results of the numerical analysis on the baryon asymmetry in the Universe.  Under the assumption that the reheating temperature is close to the maximal temperature $T_\text{max}$ at a given instant, and solving the Boltzmann equations in Eq.~(\ref{BEs}) with the initial condition on $z=M_\text{lightest}/T_\text{max}\gtrsim 0.06$, the lepton asymmetry due to the out-of-equilibrium decay of the three RH neutrinos is partially washed out by inverse decays and $\Delta L=2$ processes. 

\begin{figure}[h!]
\hspace{-5mm}
\includegraphics[width=0.45\textwidth]{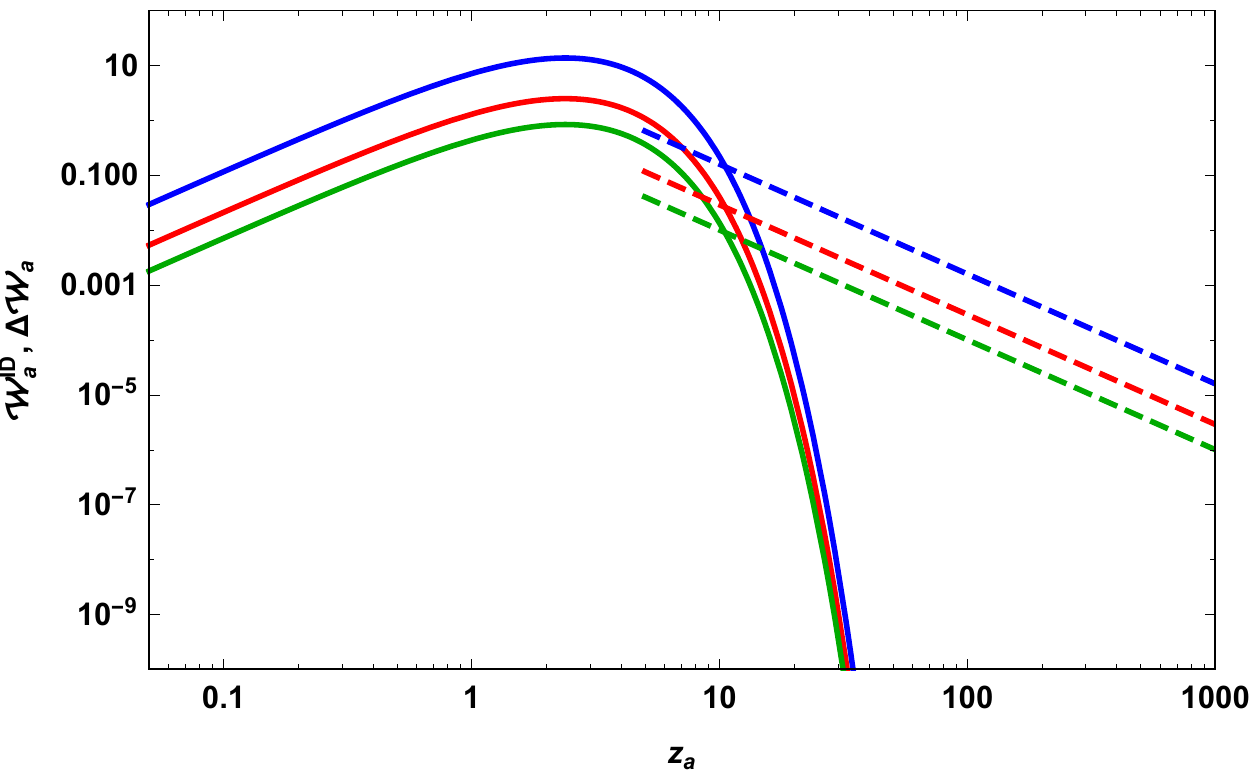}
\caption{\footnotesize\em Profiles $\W_a^{ID}$ (continuos lines) and $\Delta\W_a$ (dashed lines) as a function of $z_a$. The colours refer to the RH neutrino mass eigenstate in the NO case: the blue (red) [green] continuous line corresponds to the lightest (next-to-lightest) [heaviest] RH neutrino. The lepton number violation scale is fixed to $\mu_L=10^{16}\GeV$, the lightest active neutrino mass to $\ml=0.003\eV$, which corresponds to $z_\text{in}=0.06$, and the coefficients $c_1=c_2=0.01$.}
\label{Fig.Washout}
\end{figure}

Fig.~\ref{Fig.Washout} shows the profiles of $\W_a^{ID}$ (continuos lines) and $\Delta\W_a$ (dashed lines) as a function of $z_a$: the value for $z_a$ at which continuos and dashed lines cross is $z_\Delta\approx10$ and it corresponds to the temperature at which the washout due to inverse decays starts to be less relevant than the dilution effect due to the $\Delta L=2$ processes. The $\Delta\W_a$ lines start from $z_a=5$, satisfying the condition $z_a\gg1$ as discussed below Eq.~(\ref{alphaexpression}). The profiles in Fig.~\ref{Fig.Washout} correspond to a specific choice for the lepton number violation scale, $\mu_L=10^{16}\GeV$, the lightest active neutrino mass, $\ml=0.003\eV$, and the coefficients $c_1=c_2=0.01$, and it refers to the NO spectrum. Considering the IO spectrum, the main difference resides in that the lines corresponding to the lightest and the next-to-lightest neutrinos (blue and red) almost overlap. Lowering $\mu_L$, taking larger values for $\ml$ or taking different values for $c_{1,2}$, but still smaller than $0.1$, does not change substantially the plot. Instead, for values $c_{1,2}\sim1$, the washout effects of the heaviest neutrino become more relevant, although not changing the global picture. It follows from the fact that so large $c_{1,2}$ values induce large off-diagonal entries in $\lambda$ in Eq.~(\ref{lambda}) and then the RH neutrino flavour directions have larger overlap.

\begin{figure}[h!]
\hspace{-5mm}
\includegraphics[width=0.45\textwidth]{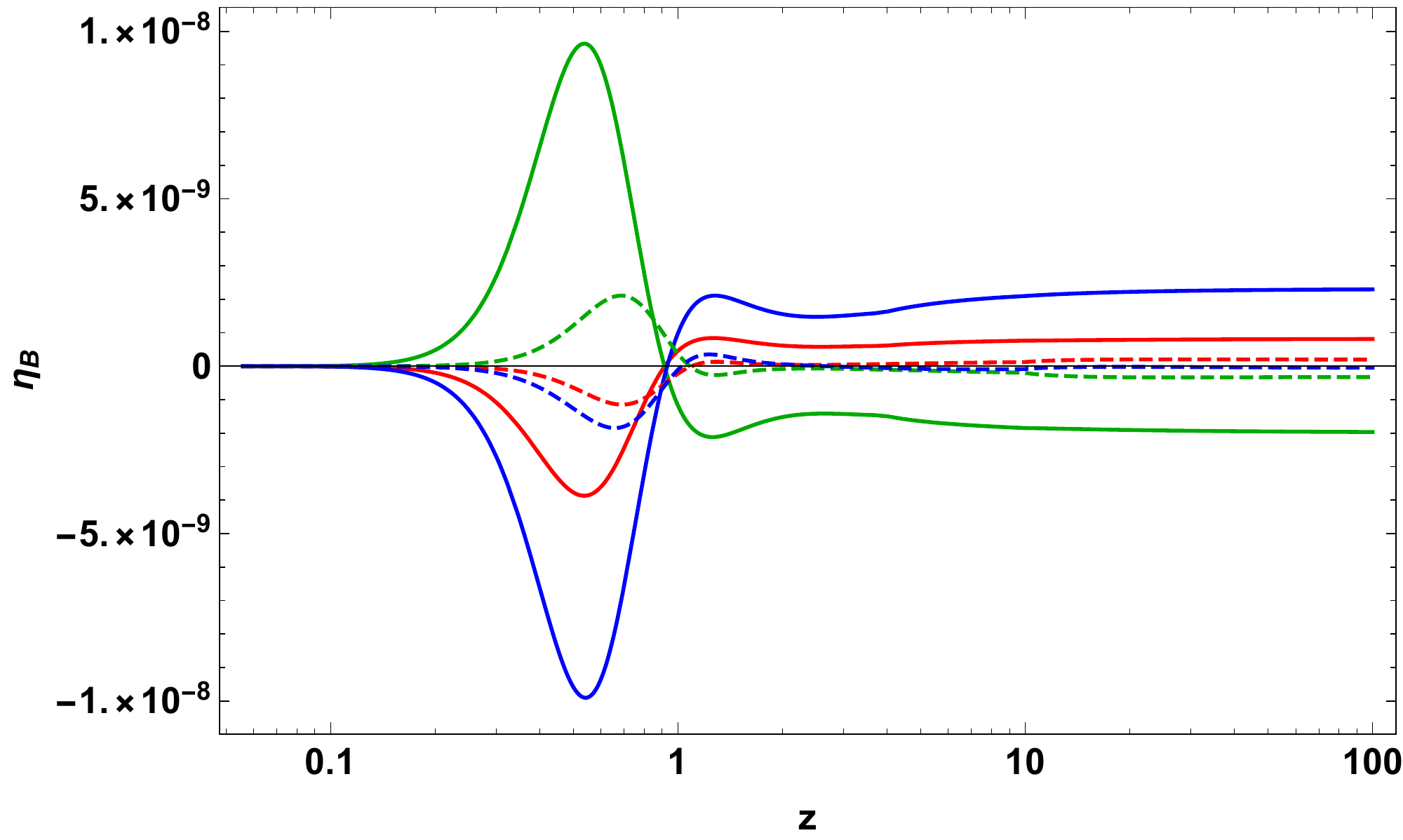}
\caption{\footnotesize\em $\eta_B$ as a function of $z\in[0.06,\,100]$ for three benchmark points in the parameter space: the green line corresponds to $\alpha_{21}=\pi$ and $\alpha_{31}=\pi/4$; the blue line corresponds to $\alpha_{21}=7\pi/4$ and $\alpha_{31}=\pi/2$;  the red line  to $\alpha_{21}=3\pi/4$ and $\alpha_{31}=5\pi/4$. Continuous (dashed) lines correspond to the NO (IO) case. The mass of the lightest active neutrino is fixed to $\ml=0.02\eV$, while the remaining input parameters have been taken at their central values as reported in Tab.~\ref{TableOscFit}. }
\label{Fig.EtaB_z_max}
\end{figure}

The standard procedure consists in solving the Boltzmann equations with a final value $z_a=+\infty$, even if this not effective from a computational point a view. However, it is possible to identify a value $z_\text{max}$ such that $\eta_B$ is practically constant for $z_a>z_\text{max}$.  The profile of $\eta_B$ as a function of $z_a$ is shown in Fig.~\ref{Fig.EtaB_z_max} for three distinct benchmark points in the parameter space: in a good approximation $z_\text{max}=20$ and this value will be adopted in the rest of the analysis.

\begin{figure}[h!]
\centering
\subfloat[$\eta_B$ {\it vs} $\ml$ for the NO case]
{\includegraphics[width=0.45\textwidth]{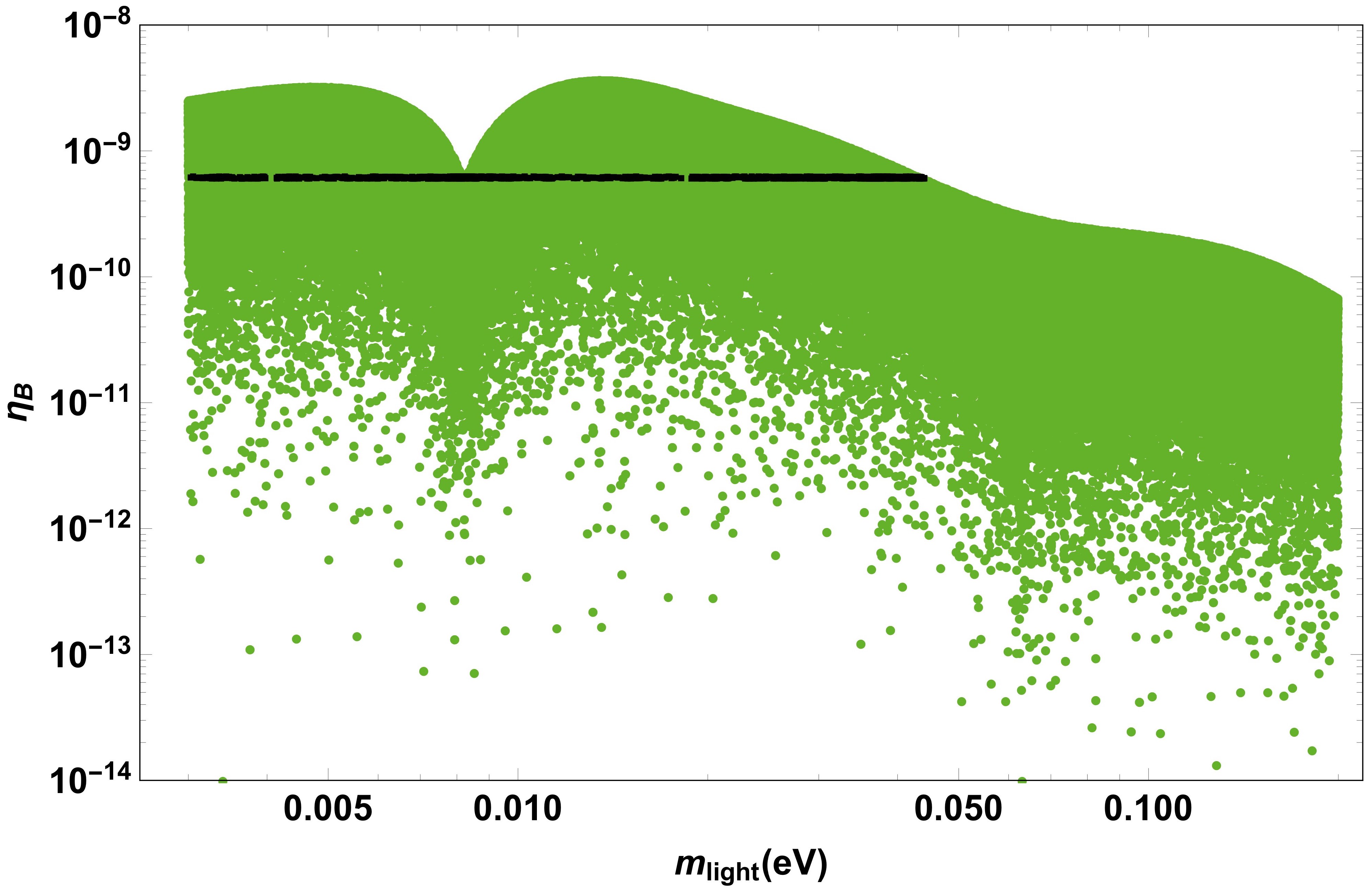}}
\hfil
\subfloat[$\eta_B$ {\it vs} $\ml$ for the IO case]
{\includegraphics[width=0.45\textwidth]{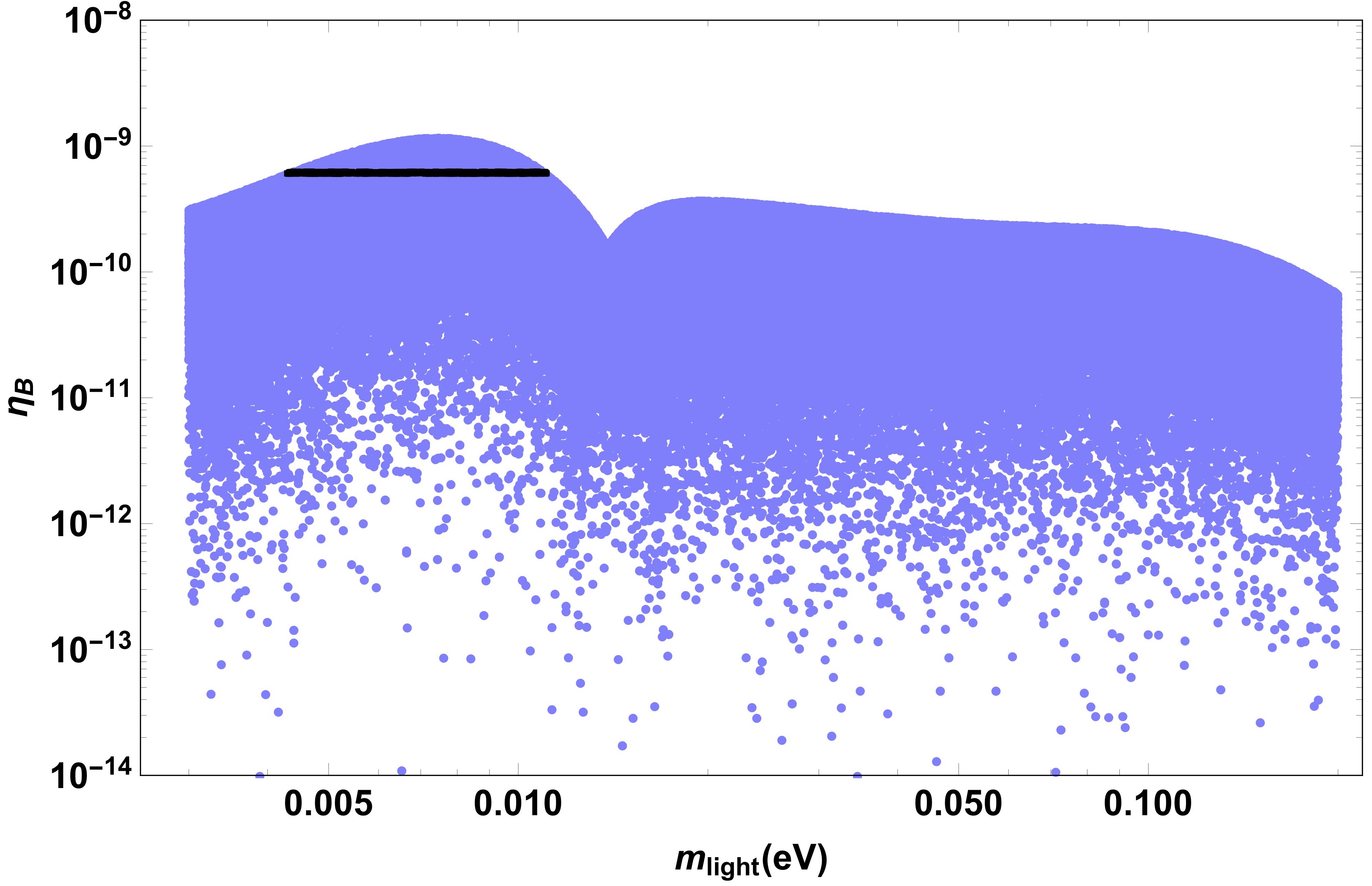}}
\caption{\footnotesize\em $\eta_B$ as a function of the lightest neutrino mass for the NO on the top and IO on the bottom. In black the points where $\eta_B$ falls inside its experimental determination at $3\sigma$ error. Charged lepton masses and neutrino oscillation parameters have been taken at their central value as in Tab.~\ref{TableOscFit}, $0.01\lesssim z<20$, $c_1=c_2=0.01$ and the Majorana CP phases randomly vary in their dominium.}
\label{fig:EtaB}
\end{figure}

Moreover, Fig.~\ref{Fig.EtaB_z_max} leads to the conclusion that $\eta_B$ strongly depends on the specific benchmark point chosen and in consequence one may expect that only a small percentage of points in the whole parameter space accommodates the current determination of $\eta_B$. This is reflected in the scatter plots in Fig.~\ref{fig:EtaB} that show $\eta_B$ as a function of the lightest active neutrino mass, for $c_1=c_2=0.01$ (details on the input parameters can be found in the caption):  values for $\eta_B$ consistent with data, represented by the black points in the plots, can be found for $m_\text{lightest}\in[0.003,\,0.04]\eV$ in the NO case and for $m_\text{lightest}\in[0.004,\,0.012]\eV$ in the IO case. $\eta_B$ cannot take values in the white region above the coloured ones, while any arbitrary smaller value is not excluded, although much smaller ones would correspond to fine-tuned situations where cancellations between the final contributions to $\eta_B$ occur.

The cuspids at $\ml\sim0.008\eV$ in the NO and at $\ml\sim0.012\eV$ in the IO do not correspond to any cancellation in the $\varepsilon_{\alpha\beta}$ parameters, but they arise as a numerical output during the resolution of the Boltzmann equations.

\begin{figure*}
\centering
\subfloat[$\alpha_{21}$ {\it vs} $\ml$ for the NO case\label{fig:CorrelationMajoranaa}]
{\includegraphics[width=0.45\textwidth]{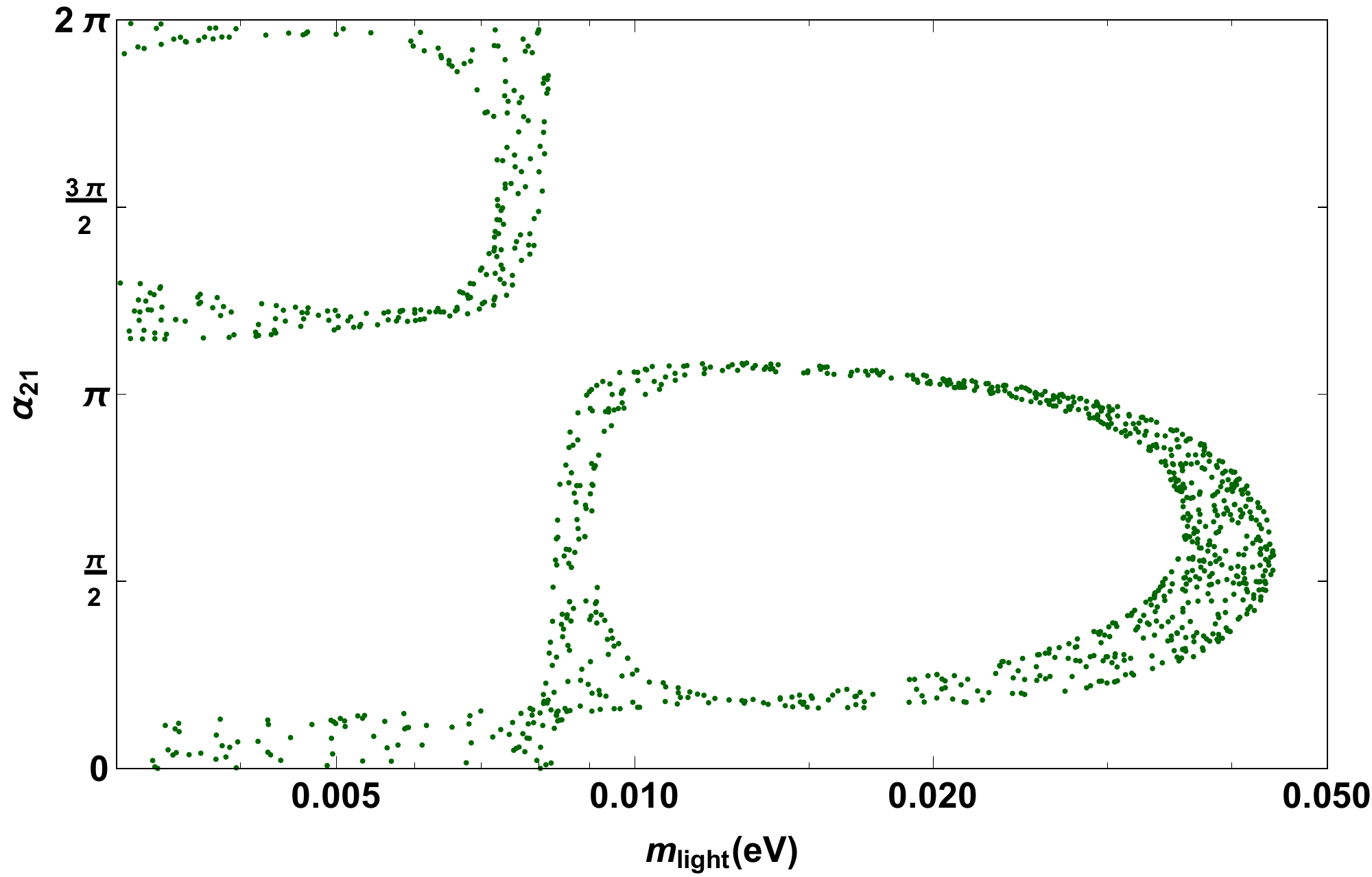}}
\hfil
\subfloat[$\alpha_{21}$ {\it vs} $\ml$ for the IO case \label{fig:CorrelationMajoranab}]
{\includegraphics[width=0.45\textwidth]{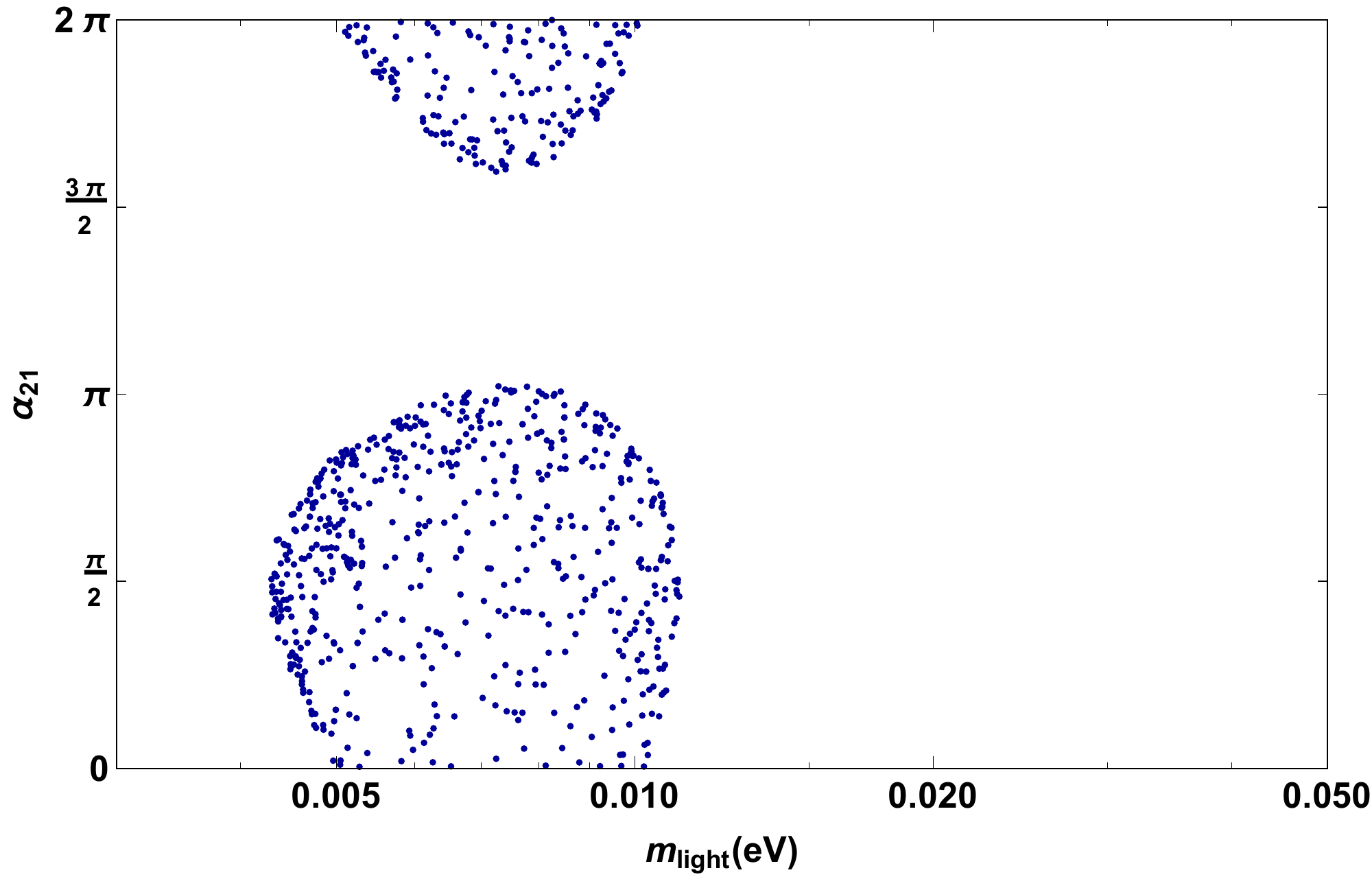}}\\
\subfloat[$\alpha_{31}$ {\it vs} $\ml$ for the IO case \label{fig:CorrelationMajoranac}]
{\includegraphics[width=0.45\textwidth]{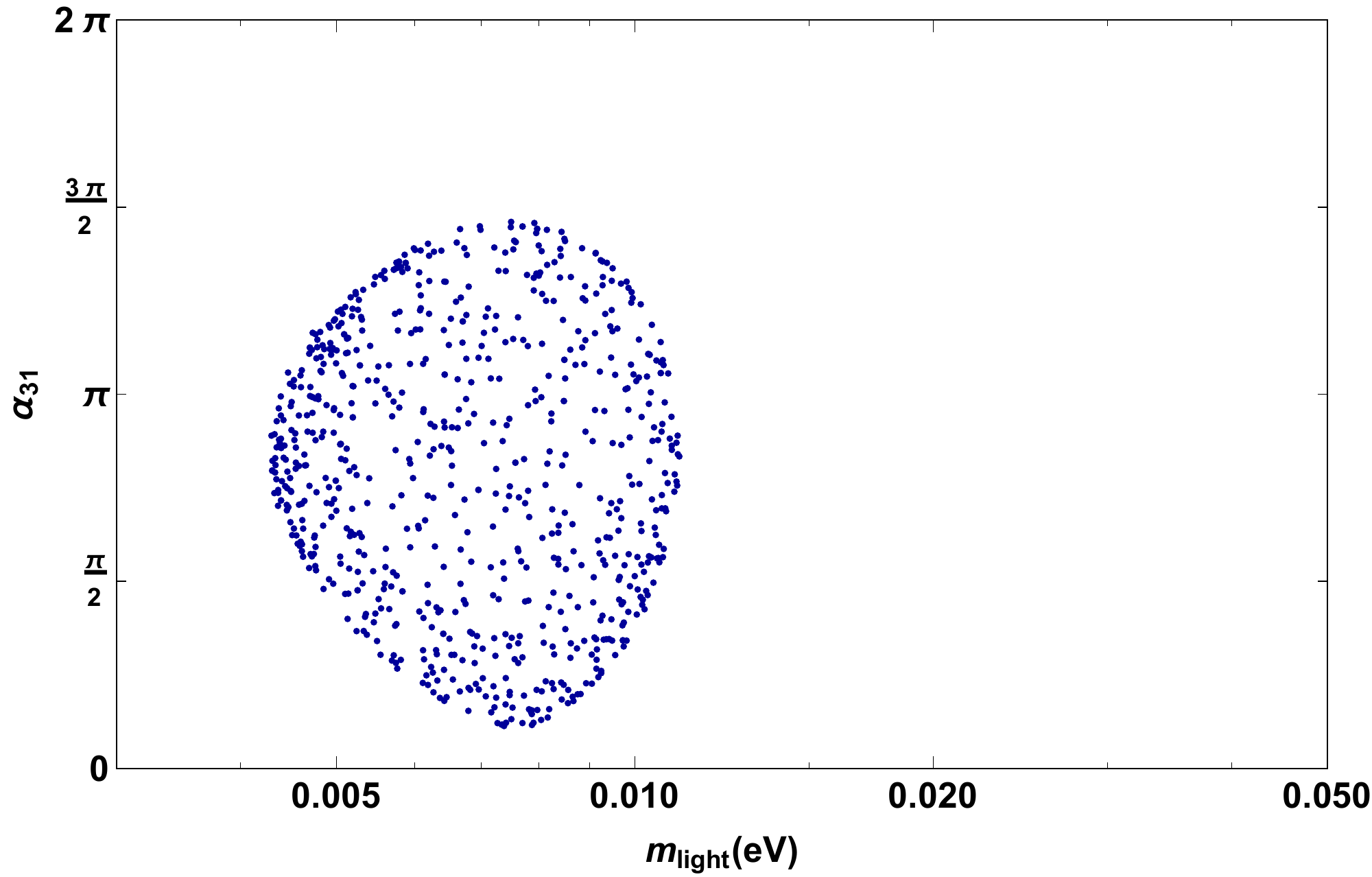}}
\hfil
\subfloat[$\alpha_{31}$ {\it vs} $\alpha_{21}$ for the IO case \label{fig:CorrelationMajoranad}]
{\includegraphics[width=0.45\textwidth]{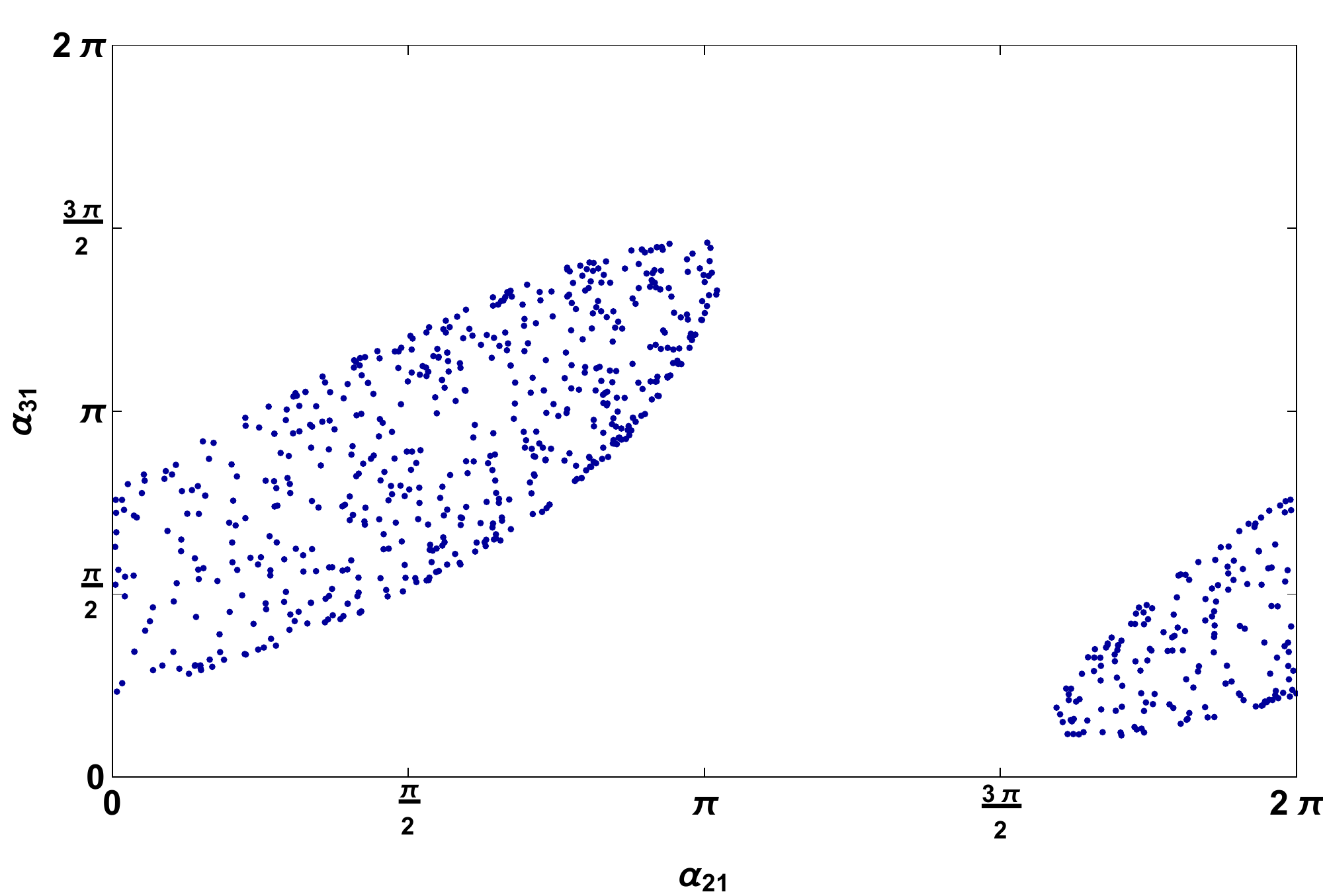}}
\caption{\footnotesize\em Correlation plots for the Majorana CP phases and the lightest active neutrino mass for the only points that satisfy $\eta_B$ within its experimental determination at $3\sigma$ error. } 
\label{fig:CorrelationMajorana}
\end{figure*}

Fig.~\ref{fig:CorrelationMajorana} shows the correlations existing between the Majorana CP phases and the lightest active neutrino mass for the NO case in \ref{fig:CorrelationMajoranaa} and for the IO in \ref{fig:CorrelationMajoranab} and \ref{fig:CorrelationMajoranac}, and between the two Majorana phases for the only IO case in \ref{fig:CorrelationMajoranad}. The $\alpha_{31}$ phase does not manifest any relevant correlation for the NO case. The plots suggest the presence of specific regions of the parameter space corresponding to a successful baryogenesis. For the NO case, one may conclude that $\alpha_{21}$ and $\ml$ are highly correlated and, for a given value of $\ml$, $\alpha_{21}$ varies only inside a small interval. This is not the case for the Majorana phases in the IO case, where the allowed parameter space is much wider; however, the strong correlation between them in Fig.~\ref{fig:CorrelationMajoranad} identifies specific regions of values where $\eta_B$ agrees with data at $3\sigma$.\\

The scatter plots shown in Figs.~\ref{fig:EtaB} and \ref{fig:CorrelationMajorana} are obtained with the Dirac CP phase within its $1\sigma$ confidence level, that nowadays is a large interval of $\sim60^\circ$ and $\sim80^\circ$ for the NO and IO respectively. These results have a very mild dependence on the value of this phase: by comparing the specific predictions for distinct fixed values of $\delta^\ell_{CP}$, no relevant differences can be appreciated. 

On the other hand, these plots highly depend on the values of $c_1=c_2$: for smaller values, for example $c_1=c_2=0.001$, $\eta_B$ is predicted to be smaller than its experimental determination at $3\sigma$ in the whole range for $\ml$ and for both NO and IO; for larger values, for example $c_1=c_2=0.1$, points with $\eta_B=6\times 10^{-10}$ can be found for any value of $\ml$ and in both NO and IO, but no correlation between Majorana phases and $\ml$ are present. In the latter case, a successful description of BAU is the result of an occasional cancellation between the contributions to $\eta_B$ obtained solving the density matrix equations in Eq.~(\ref{BEs}).

\begin{figure*}
\centering
\subfloat[Normal Hierarchical case]
{\includegraphics[width=0.32\textwidth]{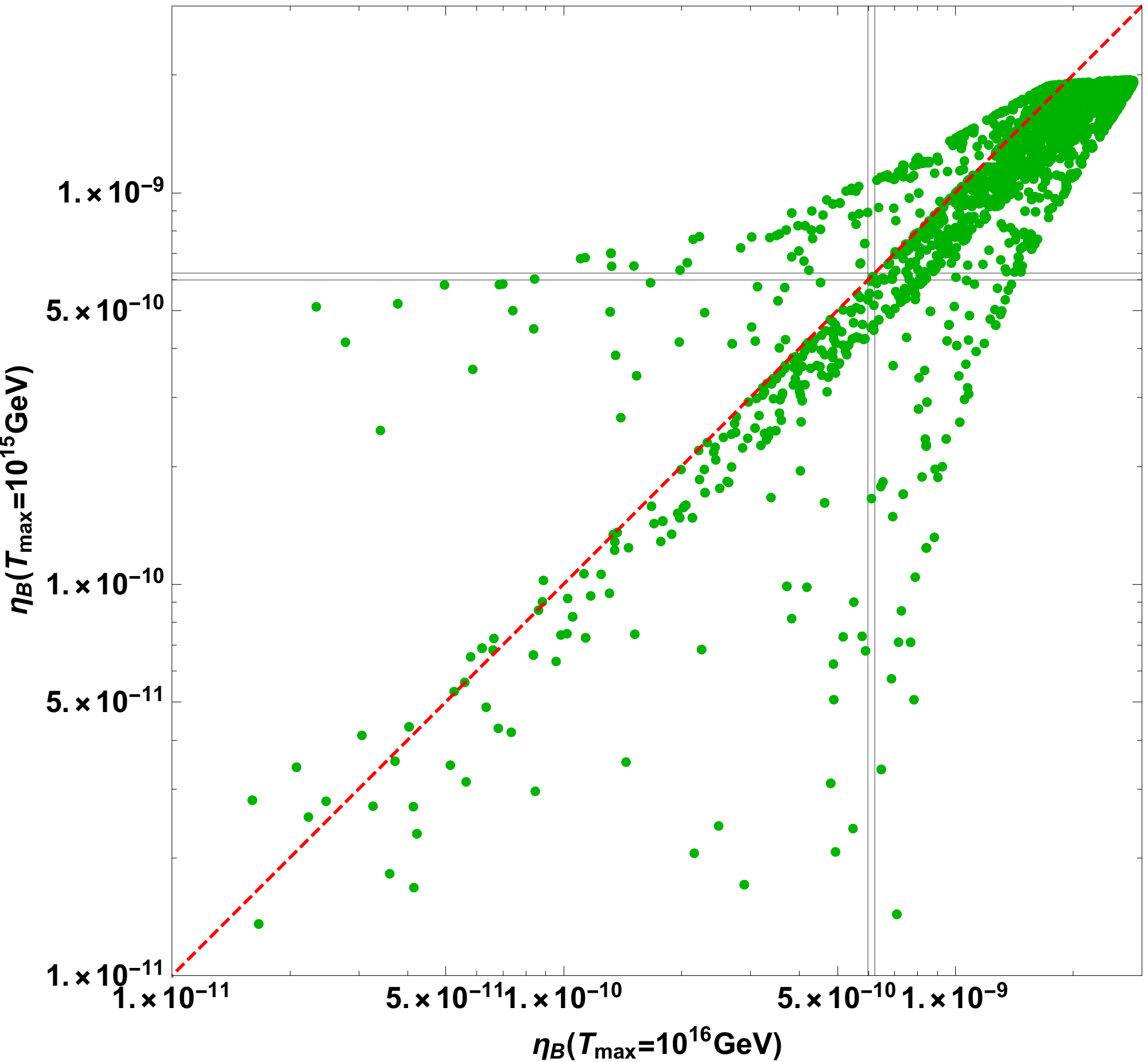}}
\hfil
\subfloat[Inverse Hierarchical case]
{\includegraphics[width=0.32\textwidth]{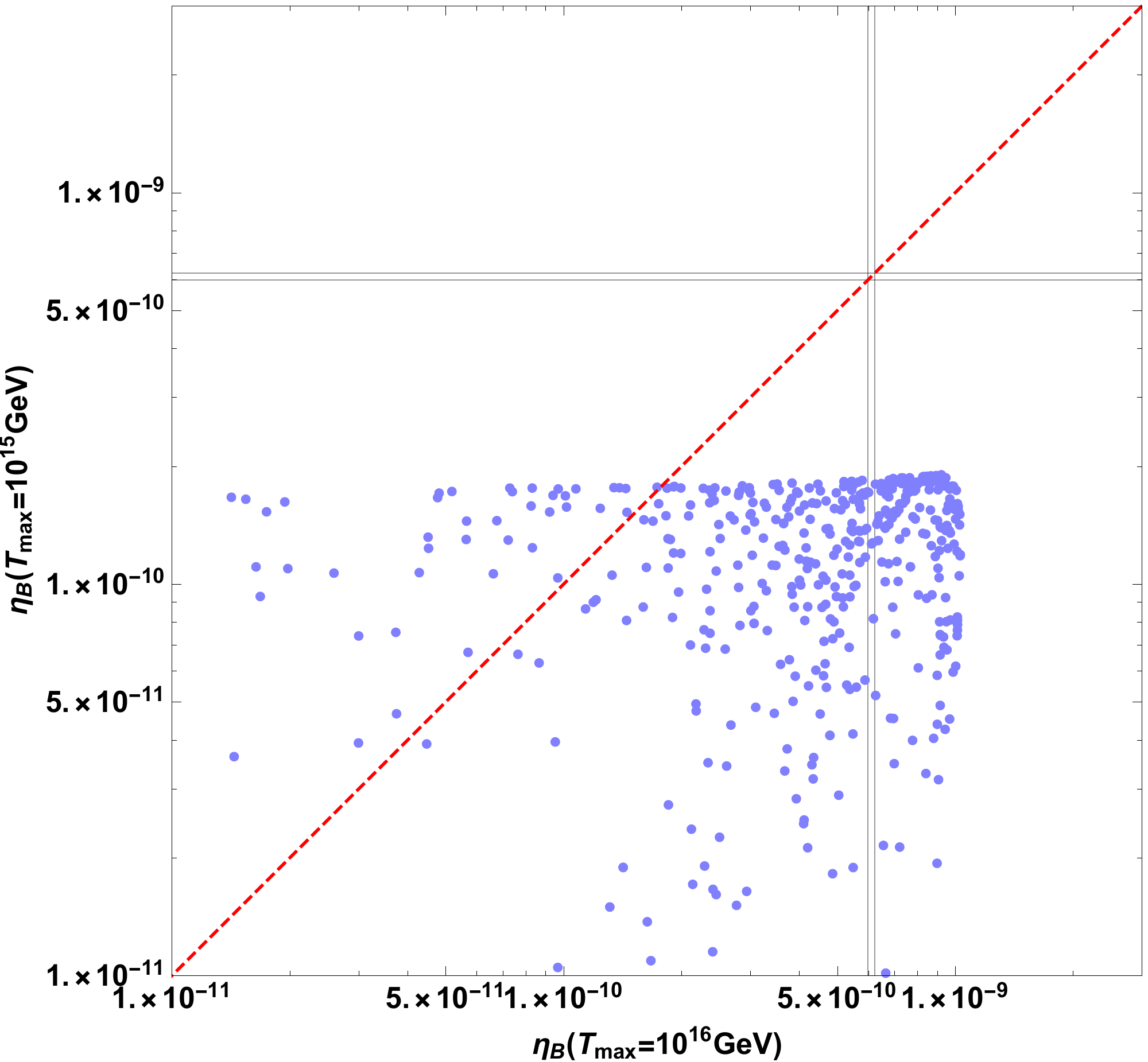}}
\hfil
\subfloat[Degenerate case]
{\includegraphics[width=0.32\textwidth]{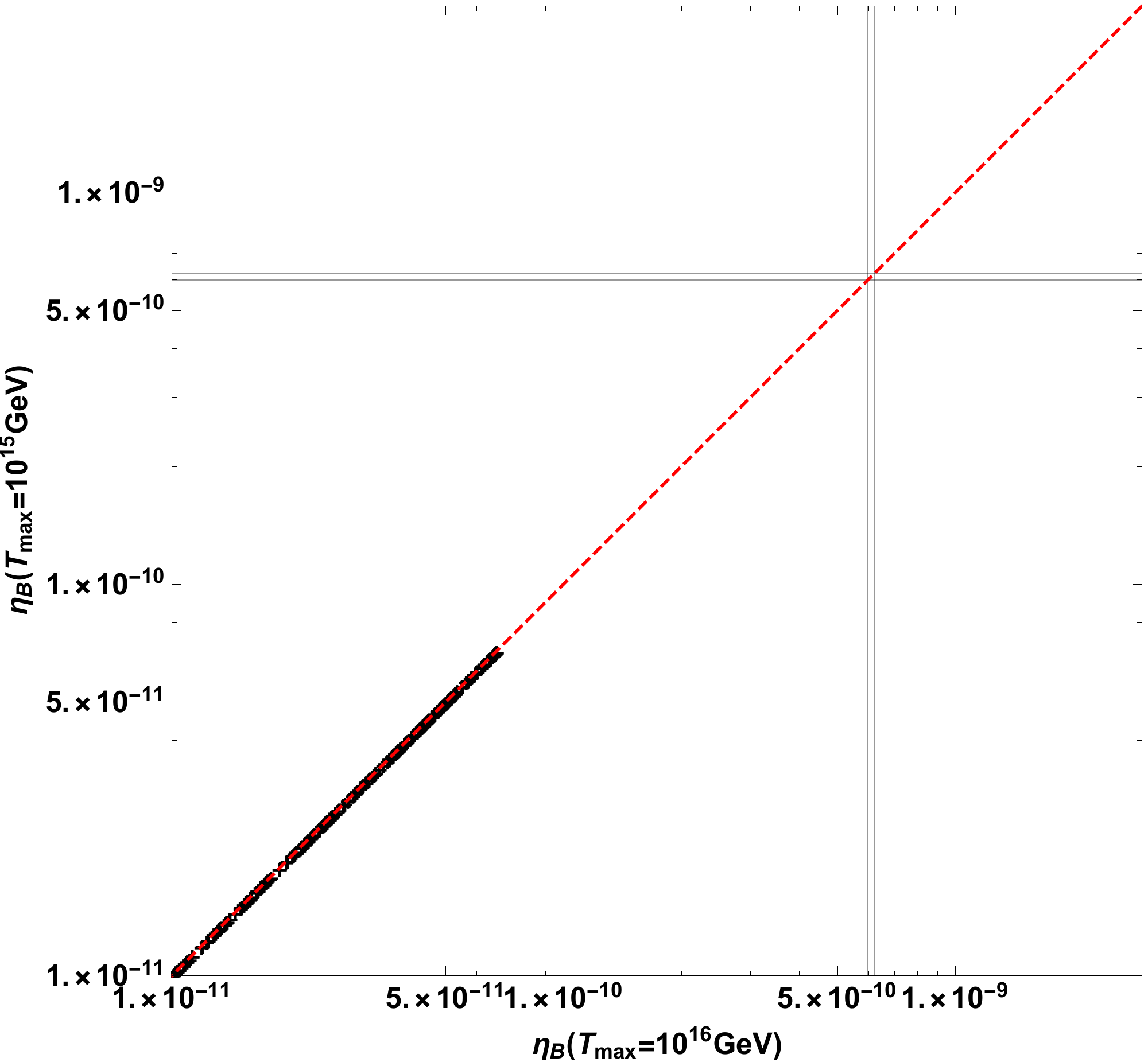}}
\caption{\footnotesize\em Correlation plots of $\eta_B$ with different $T_\text{max}$. In the axis of the abscissas $\eta_B$ with $T_\text{max}=10^{16}\GeV$, while in the one of the ordinates $\eta_B$ with $T_\text{max}=10^{15}\GeV$. On the left the normal hierarchical case, in the center the inverse hierarchical one, and on the right the degenerate spectrum case. The red dashed line represents the diagonal to easier drive the eye on the values when the two computed $\eta_B$ have the same value. The black continuous lines delimit the $3\sigma$ value for the experimental determination of $\eta_B$. The two parameters $c_1$ and $c_2$ have been fixed at $0.01$, while two values for $\ml$ have been considered, $\ml=0.006\eV$ for the first two plots and $\ml=0.2\eV$ for the one on the right. Each point in the plots corresponds to a given random choice of the rest of parameters.} 
\label{fig:CorrelationEtaB}
\end{figure*}

The subjacent hypothesis to the numerical result shown above is that the maximal temperature of the Universe is $T_\text{max}=\mu_L=10^{16}\GeV$, implying that the three RH neutrinos are thermally produced and contribute to the final value of $\eta_B$. If a lower value for $T_\text{max}$ is taken, then the heaviest neutrinos may not be thermally produced and their contributions would be negligible. Fig.~\ref{fig:CorrelationEtaB} shows the effect on the final value of $\eta_B$ of lowering the value of $T_\text{max}$, for a normal hierarchical active neutrino spectrum on the left, for an inverse hierarchical one in the middle, and for a degenerate spectrum on the right. The axes represent the final value of $\eta_B$ considering $T_\text{max}=10^{16}\GeV$ and $T_\text{max}=10^{15}\GeV$. The two parameters $c_1$ and $c_2$ have been fixed at $0.01$, while two values for $\ml$ have been considered, $\ml=0.006\eV$ for the first two plots and $\ml=0.2\eV$ for the one on the right. Each point in the plots corresponds to a given random choice of the rest of parameters: in this way, it is possible to clearly identify on the final value of $\eta_B$ the impact of the temperature dependence and therefore the impact of the heavier sterile neutrinos. The diagonal red line drives the eye to tell when $\eta_B$ is larger for $T_\text{max}=10^{16}\GeV$ or for $T_\text{max}=10^{15}\GeV$: if the points align along the diagonal, then either the heaviest sterile neutrino would not contribute to the final value of $\eta_B$ or the three of them are thermally produced even considering the lowest temperature case; if all the points cover the region on the right of the diagonal, then the heaviest sterile neutrino does have an impact and its contribution sums constructively with the ones from the lightest states; in the opposite case, i.e. all the points on the left of the diagonal, its contribution sums destructively with the other ones. 

Focussing first on the normal hierarchical case (plot on the left), the points cover an area along the diagonal, with a small preference for $\eta_B$ at $T_\text{max}=10^{16}\GeV$. Any fixed value of $\eta_B$ at $T_\text{max}=10^{16}\GeV$ corresponds to the same values of $\eta_B$ at $T_\text{max}=10^{15}\GeV$, whiting a factor $2\div3$. Moreover, there are points where the $\eta_B$ matches with the experimentally allowed regiones (inside the parallel continuous black lines) and many others where this does not occurs. This lets conclude that the value of $\eta_B$ strongly depends on the specific set of parameters, especially Majorana phases, considered, as already pointed out in Fig.~\ref{fig:EtaB}. Moreover, the value for $\eta_B$ with $T_\text{max}=10^{16}\GeV$, where all the three sterile neutrinos contribute, are within a factor $2\div3$ similar to the ones for  $\eta_B$ with $T_\text{max}=10^{15}\GeV$, where only the lightest ones are relevant. The small preference for the region where $\eta_B$ with $T_\text{max}=10^{16}\GeV$ indicates that the impact of the heaviest sterile neutrino is often not negligible and slightly increases the final value of $\eta_B$. It follows that Fig.~\ref{fig:EtaB}(a), where the points show that $\eta_B$ spans a few order of magnitudes, is a good representative for this scenario with $T_\text{max}=10^{16\div15}\GeV$ and for a hierarchical spectrum.

For the inverse hierarchical case (plot in the middle), the largest majority of the points cover the region for $\eta_B$ with $T_\text{max}=10^{16}\GeV$, indicating that the heaviest sterile neutrino typically contributes to the final value of $\eta_B$, increasing its value. Moreover, only for $T_\text{max}=10^{16}\GeV$, $\eta_B$ reaches the experimentally allowed region, indicating that the heaviest sterile neutrino contributions are necessary. As a result, Fig.~\ref{fig:EtaB}(b) fairly represents only the case with $T_\text{max}=10^{16}\GeV$.

Finally, focussing to the degenerate spectrum (plot on the right), all the points strictly align with the diagonal, indicating that $\eta_B$ does not change for $T_\text{max}=10^{15}\GeV$ or $10^{16}\GeV$. This was expected because for $\ml=0.2\eV$ all the three sterile neutrinos have masses below $T_\text{max}=10^{15}\GeV$ and therefore are the three of them thermally generated. Both the plots in Fig.~\ref{fig:EtaB} well represent this scenario with $T_\text{max}=10^{16\div15}\GeV$ for the degenerate spectrum.

The plots equivalent to those in Figs.~\ref{fig:EtaB} and \ref{fig:CorrelationMajorana}(a) for $T_\text{max}=10^{15}\GeV$ can be found in App.~\ref{APPENDIX}. As can be seen, the NO case is essentially unaffected by the change of the temperature, while the IO one presents a difference for small values of $\ml$ where $\eta_B$ does not reach the experimental band.

\subsection{Low-energy phenomenology}

The reduction of the allowed parameter space for the Majorana phases in the $c_1=c_2=0.01$ case, Fig.~\ref{fig:CorrelationMajorana}, has an impact on the predictions for the neutrinoless double beta decay effective mass $m_{ee}$, defined by
\be
\hspace{-0.5cm}
\left|m_{ee}\right|=\left|c_{13}^2\,c_{12}^2\,m_{\nu_1}+c_{13}^2\,s_{12}^2\,m_{\nu_2}\,e^{i\alpha_{21}}+s_{13}^2\,m_{\nu_3}\,e^{i(\alpha_{31}-2\delta_\text{CP}^\ell)}\right|\,,
\ee
where $c_{ij}$ and $s_{ij}$ stand for $\cos\theta_{ij}$ and $\sin\theta_{ij}$, respectively. The investigation on this decay has received a strong impulse in the last decades and numerous experiments are currently competing to probe the existence of this process, as its observation would automatically infer that neutrinos have (at least partly) Majorana nature~\cite{Schechter:1981bd}. Tab.~\ref{tab:0nu2betaExp} reports the lower bounds on $|m_{ee}|$ sensitivity for near future $0\nu2\beta$ experiments that will be considered in the following.

\begin{table}[h!]
\begin{center}
\begin{tabular}{l|c|c|}
Experiment						& Isotope			& $|m_{ee}|\,[\eV]$ \\[1mm]
\hline
&&\\[-2mm]
CUORE~\cite{Artusa:2014lgv} 			& ${}^{130}$Te 		& $0.073\pm0.008$ \\[1mm]
GERDA-II~\cite{Brugnera:2013xma} 		& ${}^{76}$Ge 		& $0.11\pm0.01$ \\[1mm]
LUCIFER~\cite{Pattavina:2016kqn} 		& ${}^{82}$Se 		& $0.20\pm0.02$ \\[1mm]
MAJORANA D.~\cite{Abgrall:2013rze}	& ${}^{76}$Ge 		& $0.13\pm0.01$ \\[1mm]
NEXT~\cite{Laing:2016puy}			& ${}^{136}$Xe 	& $0.12\pm0.01$ \\[1mm]
AMoRE~\cite{Jo:2017jod}				& ${}^{100}$Mo 	& $0.084\pm0.008$ \\[1mm]
nEXO~\cite{Pocar:2015ota}			& ${}^{136}$Xe 	& $0.011\pm0.001$ \\[1mm]
PandaX-III~\cite{Chen:2016qcd}		& ${}^{136}$Xe 	& $0.082\pm0.009$ \\[1mm]
SNO+~\cite{Andringa:2015tza}			& ${}^{130}$Te 		& $0.076\pm0.007$ \\[1mm]
SuperNEMO~\cite{Arnold:2015wpy} 		& ${}^{82}$Se		& $0.084\pm0.008$ \\[1mm]
\hline
\end{tabular}
\end{center}
\caption{\footnotesize\it Lower bounds for $|m_{ee}|$ for the next future sensitivities and/or experiments on $0\nu2\beta$ decay.}
\label{tab:0nu2betaExp}
\end{table}

\begin{figure*}
\centering
\subfloat[$|m_{ee}|$ {\it vs} $\ml$ for the NO case \label{fig:LowEnergyMeea}]
{\includegraphics[width=0.45\textwidth]{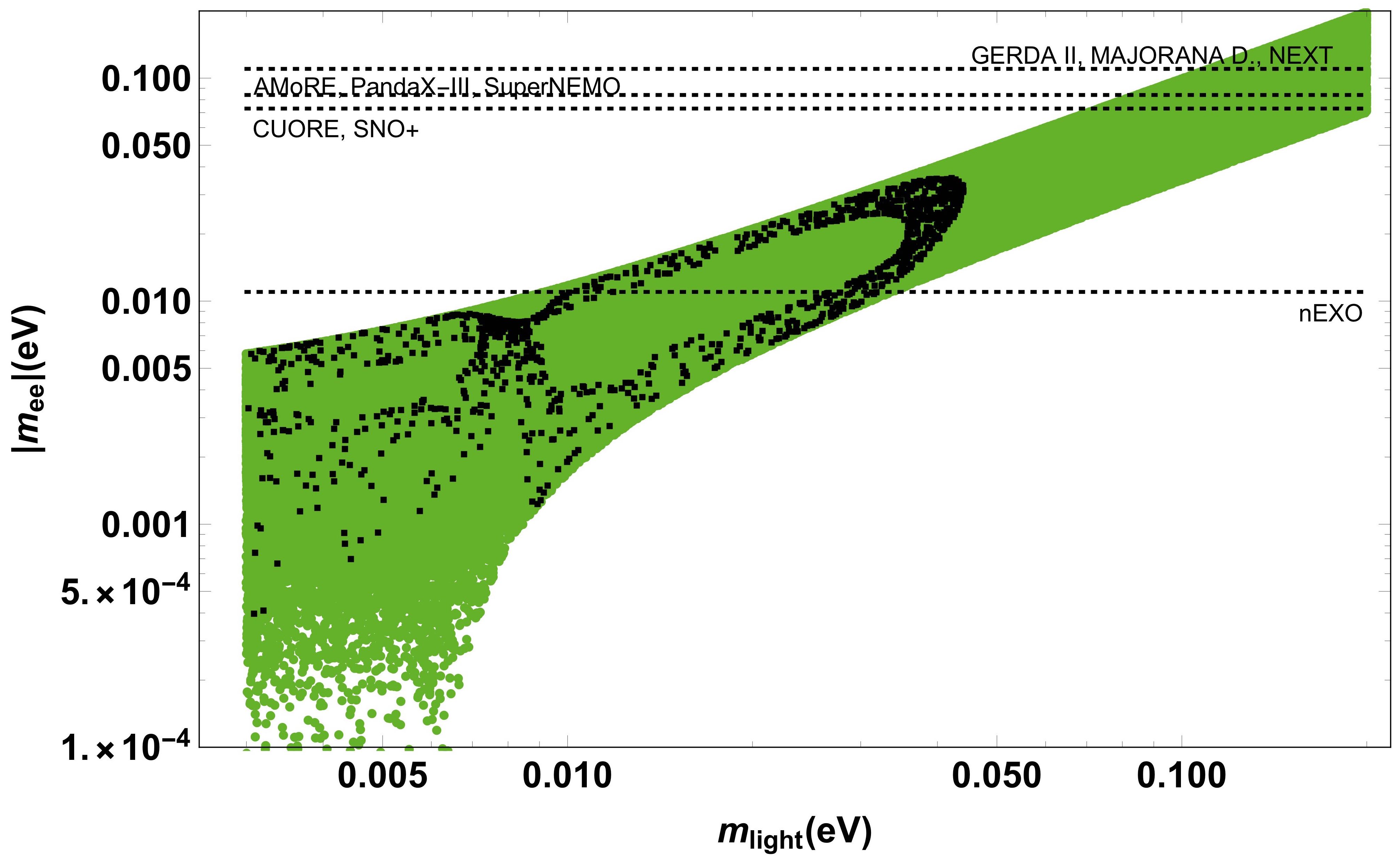}}
\hfil
\subfloat[$|m_{ee}|$ {\it vs} $\ml$ for the IO case \label{fig:LowEnergyMeeb}]
{\includegraphics[width=0.45\textwidth]{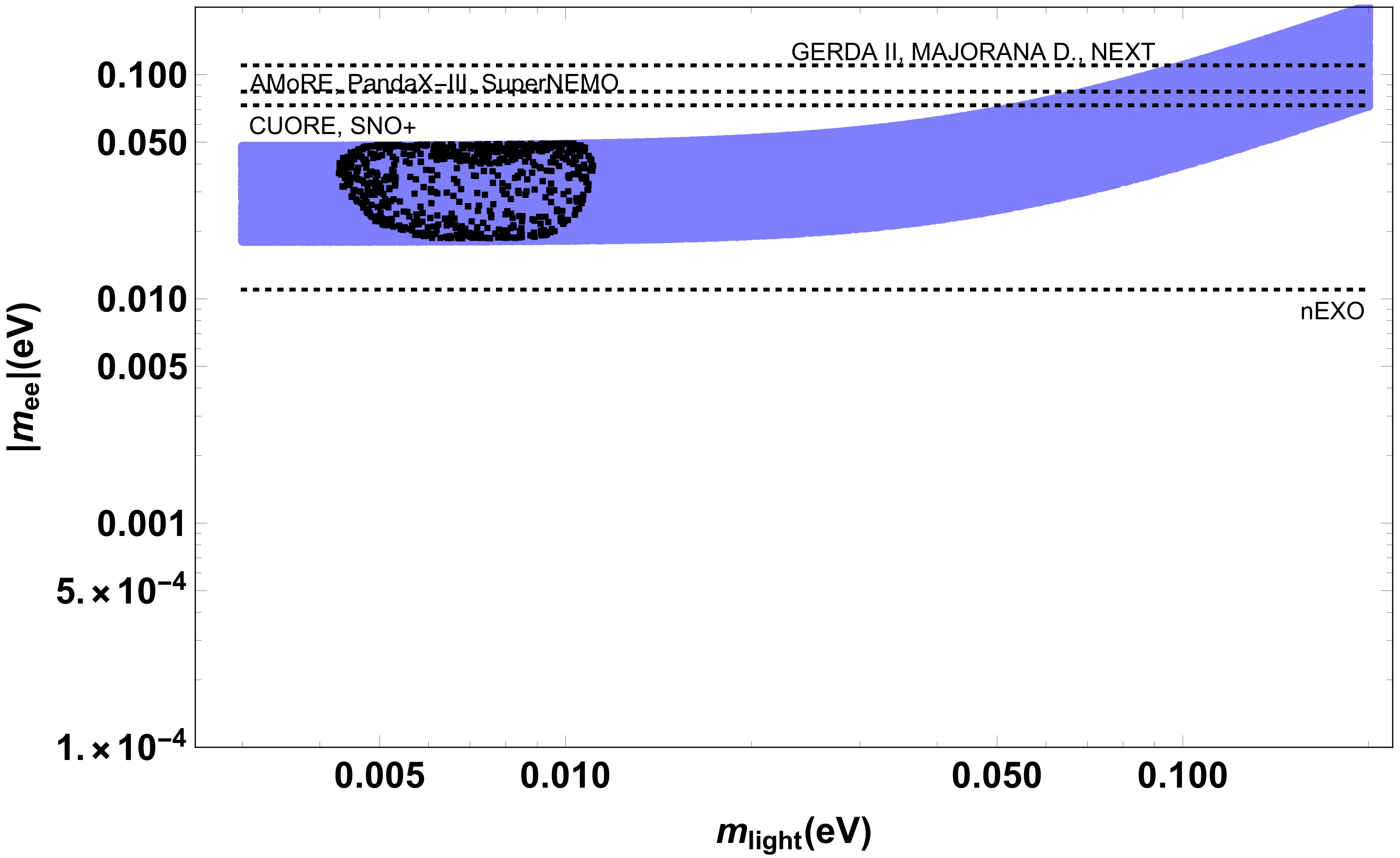}}\\
\subfloat[$|m_{ee}|$ {\it vs} $\alpha_{21}$ for the NO case \label{fig:LowEnergyMeec}]
{\includegraphics[width=0.45\textwidth]{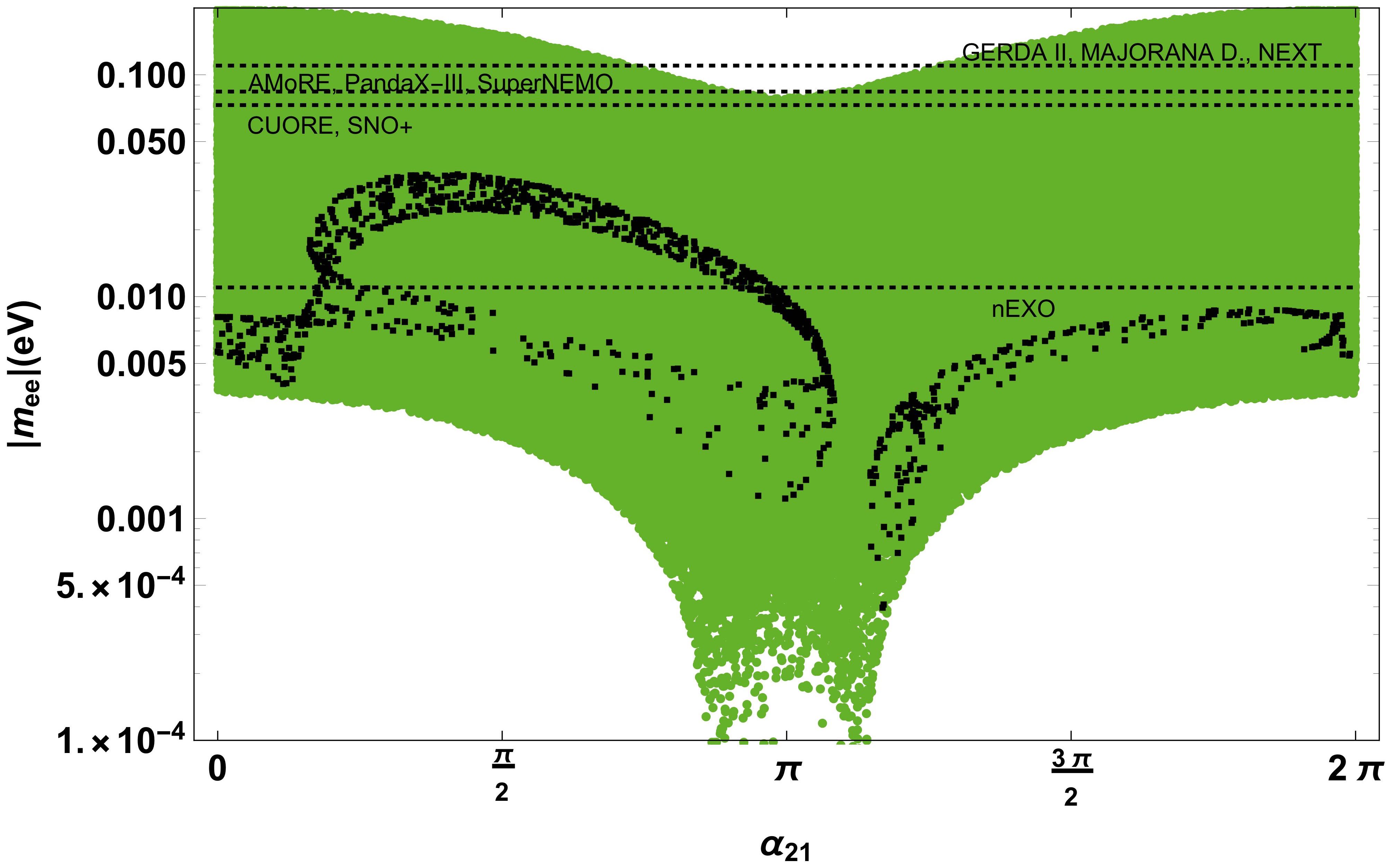}}\\
\subfloat[$|m_{ee}|$ {\it vs} $\alpha_{21}$ for the IO case \label{fig:LowEnergyMeed}]
{\includegraphics[width=0.45\textwidth]{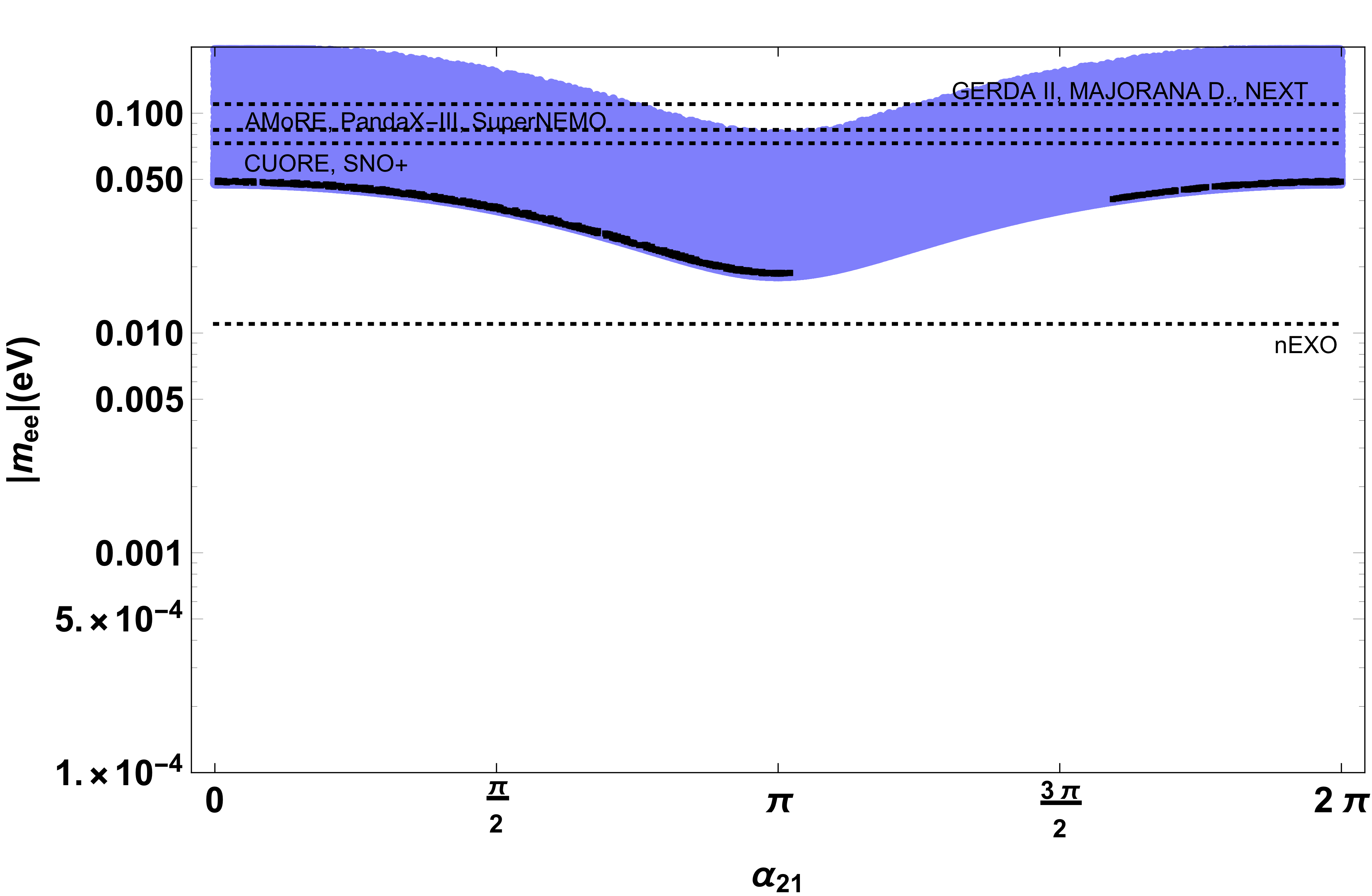}}
\hfil
\subfloat[$|m_{ee}|$ {\it vs} $\alpha_{31}$ for the IO case \label{fig:LowEnergyMeee}]
{\includegraphics[width=0.45\textwidth]{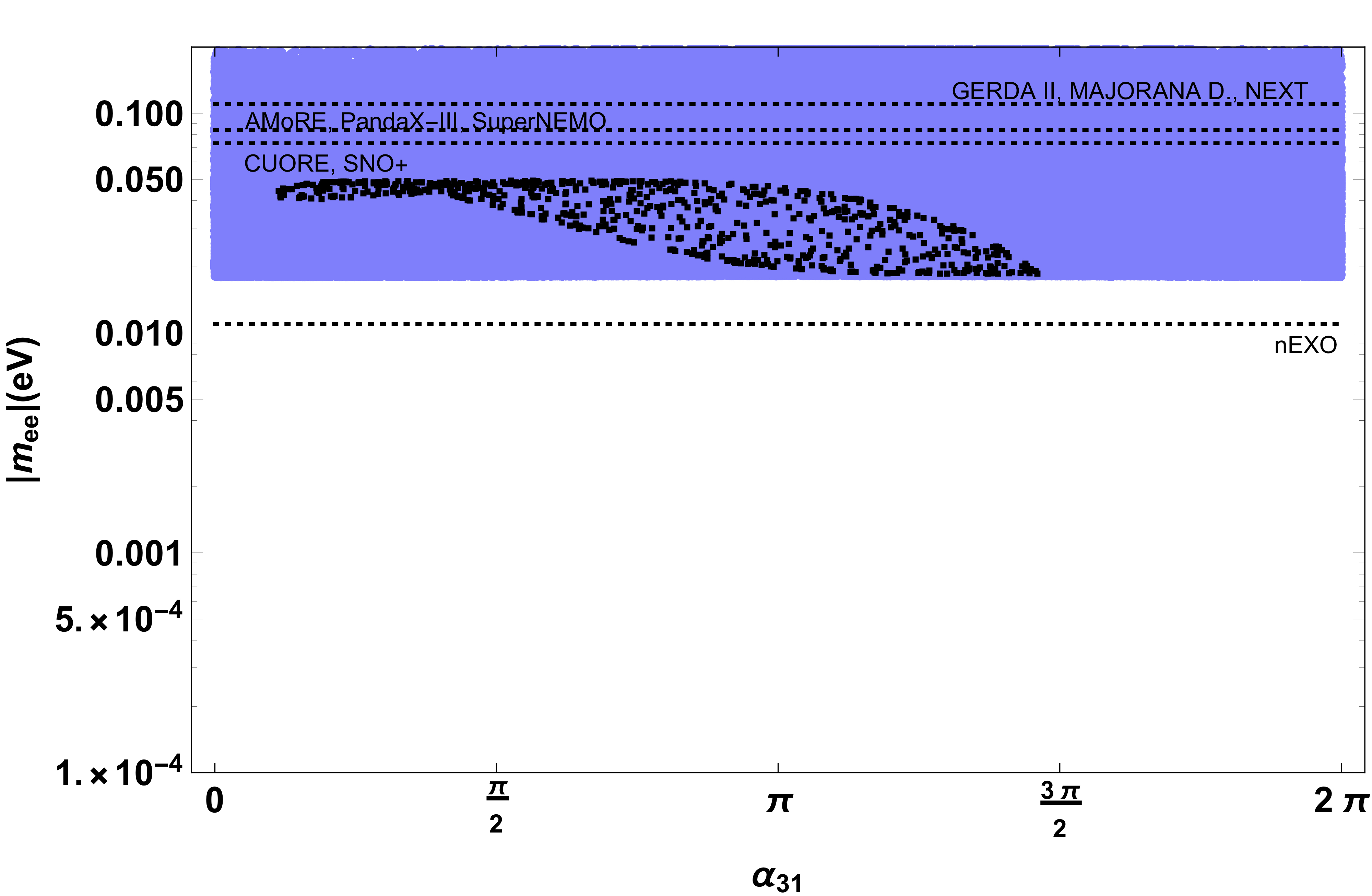}}
\caption{\footnotesize\em $|m_{ee}|$ as a function of $\ml$ in a) for the NO case and in b) for the IO one. $|m_{ee}|$ as a function of the Majorana phases for the NO case in c) and for the IO one in d) and e).}
\label{fig:LowEnergyMee}
\end{figure*}

Fig.~\ref{fig:LowEnergyMee} shows the profile of $|m_{ee}|$ as a function of the lightest active neutrino mass $\ml$ in \ref{fig:LowEnergyMeea} for the NO and in \ref{fig:LowEnergyMeeb} for the IO, while as a function of the Majorana phases in \ref{fig:LowEnergyMeec} for the NO and in \ref{fig:LowEnergyMeed} and \ref{fig:LowEnergyMeee} for the IO. For both the mass orderings, describing successfully the amount of BAU leaves viable only the hierarchical regime. For the NO, Fig.~\ref{fig:LowEnergyMeea}, $|m_{ee}|$ can take values only below $0.04\eV$, while a lower bound at about $4\times 10^{-4}\eV$ seems plausible, as confirmed in Fig.~\ref{fig:LowEnergyMeec}, although the point density is poor in this region: interestingly, it appears a region precluded for $0.0095\eV\lesssim\ml\lesssim0.035\eV$. For the IO, Fig.~\ref{fig:LowEnergyMeeb}, the parameter space corresponding to $\eta_B$ inside its experimental determination at $3\sigma$ is confined in a well-defined region between $0.005\eV\lesssim\ml\lesssim0.01\eV$ and $0.018\eV\lesssim|m_{ee}|\lesssim0.05\eV$. 

Complementary information can be extracted in the plots with $|m_{ee}|$ as a function of the Majorana phases. For the NO, Fig.~\ref{fig:LowEnergyMeec}, only $|m_{ee}|$ {\em vs} $\alpha_{21}$ shows a correlation: only values for $\alpha_{21}$ in the interval $~[\pi/8,\,3\pi/4]$ leads to larger values of $|m_{ee}|$, while smaller values may be described for almost any $\alpha_{21}$. For the IO, Figs.~\ref{fig:LowEnergyMeed} and \ref{fig:LowEnergyMeee}, a correlation between $|m_{ee}|$ and both the Majorana phases is present and the allowed parameter space is limited in relatively small regions.

An observation of the neutrinoless double beta decay in the present experiments, if fully interpreted in terms of Majorana neutrino exchange, would be crucial to determine the values of the Majorna phases for which a successful BAU occurs. Once determined the ordering of the active neutrino mass spectrum, a larger value for $|m_{ee}|$ would favour values of $\alpha_{21}$ in the interval $\sim[\pi/8,\,3\pi/4]$ for the NO and $\sim[-\pi/2,\,\pi/2]$ in the IO, and values of $\alpha_{31}$ in the interval $\sim[\pi/8,\,\pi]$ in the only IO. The determination of the value for the lightest active neutrino mass would help reducing these interval: if $\ml$ is found relatively large, then only the NO scenario would be compatible with a successful explanation of the BAU, while the IO case would be then excluded.

%
%
\section{Conclusions}
\label{Sect:Conc}

The MFV ansatz works extraordinary well in the quark sector accommodating a huge amount of experimental measurements. If an underlying dynamics is the reason behind this hypothesis, then it is natural to expect a similar mechanism at work also in the lepton sector. Two distinct versions of the MLFV can be considered when the SM spectrum is extended by the three RH neutrinos: only if the latter transform under the same symmetry of the lepton electroweak doublets~\cite{Alonso:2013mca}, $SU(3)_{\ell_L}\times SU(3)_{N_R}\to SU(3)_V$, then violation of the CP symmetry can be described according to the recent experimental indication. 

The presence of non-vanishing CP violating phases in the leptonic mixing may be the missing ingredient in the SM to successfully describe the baryon asymmetry in the Universe. In this paper, baryogenesis through Leptogenesis has been considered for the first time within the context of the $SU(3)_V$ MLFV framework, resulting in a very predictive setup where the $\varepsilon$ parameter that describes the amount of CP violation in Leptogenesis only depends on low-energy parameters: charged lepton and active neutrino masses, PMNS parameters and two parameters of the low-energy effective description. 

Fixing the two effective parameters at their natural value $0.01$, when a baryon to photon ratio today agrees with its experimental determination at $3\sigma$ then correlations between the Majorana phases and the lightest active neutrino mass arise. The latter can be analysed considering the impact in the neutrinoless double beta decay observable: only selected regions of the whole $|m_{ee}|$ {\it vs} $\ml$ parameter space correspond to values that are consistent with a successful baryogenesis. In the NO case, only upper bounds on $|m_{ee}|$ and $\ml$ can be identified: $|m_{ee}|\lesssim0.04\eV$ and $\ml\lesssim0.04\eV$. Instead, in the IO case, $|m_{ee}|$ can take values only inside a much smaller interval $[0.02,\,0.05]\eV$ corresponding to a narrow interval for $\ml$ that is $[0.004,\,0.012]\eV$. These regions will be tested only in several years as the sensitivity required is of the order of that one expected by the nEXO experiment.

\vspace{1.3cm}
\begin{center}
\rule{5cm}{0.4pt}
\end{center}
\acknowledgments
The authors warmly thank Pasquale di Bari, Mattias Blennow, Enrique Fern\'andez Mart\'inez, Pilar Hern\'andez, Olga Mena and Nuria Rius for discussions and suggestions. They also thank the HPC-Hydra cluster at IFT. L.M. thanks the department of Physics and Astronomy of the Universit\`a degli Studi di Padova and the Fermilab Theory Division for hospitality during the writing up of the paper. L.M. acknowledges partial financial support by the Spanish MINECO through the ``Ram\'on y Cajal'' programme (RYC-2015-17173), by the European Union's Horizon 2020 research and innovation programme under the Marie Sklodowska-Curie grant agreements No 690575 and No 674896, and by the Spanish ``Agencia Estatal de Investigaci\'on'' (AEI) and the EU ``Fondo Europeo de Desarrollo Regional'' (FEDER) through the project FPA2016-78645-P, and through the Centro de excelencia Severo Ochoa Program under grant SEV-2016-0597.


\appendix

\boldmath
\section{Lowering $T_\text{max}$}
\label{APPENDIX}
\unboldmath

\begin{figure}[h!]
\centering
\subfloat[$\eta_B$ {\it vs} $\ml$ for the NO case]
{\includegraphics[width=0.45\textwidth]{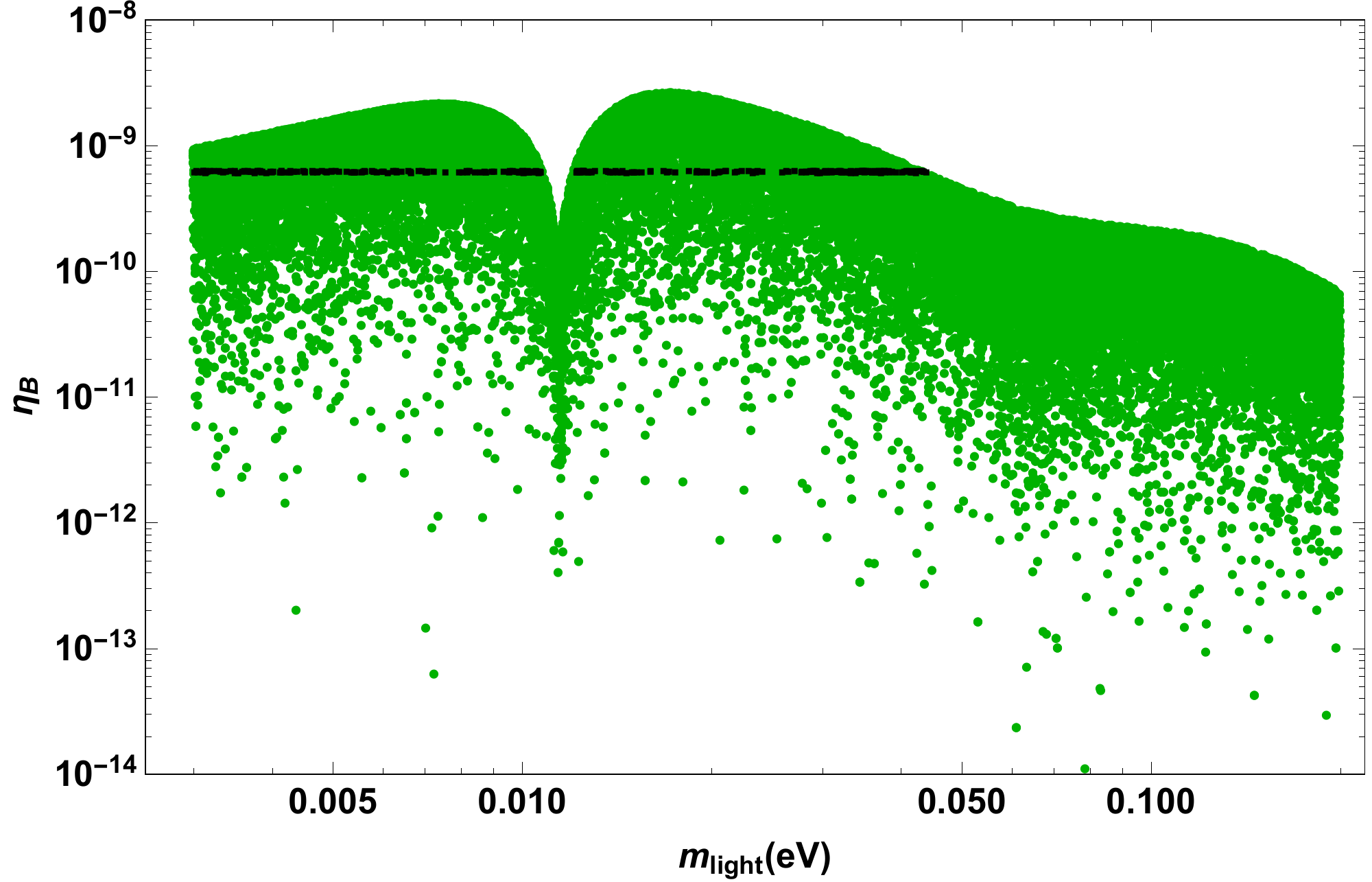}}
\hfil
\subfloat[$\eta_B$ {\it vs} $\ml$ for the IO case]
{\includegraphics[width=0.45\textwidth]{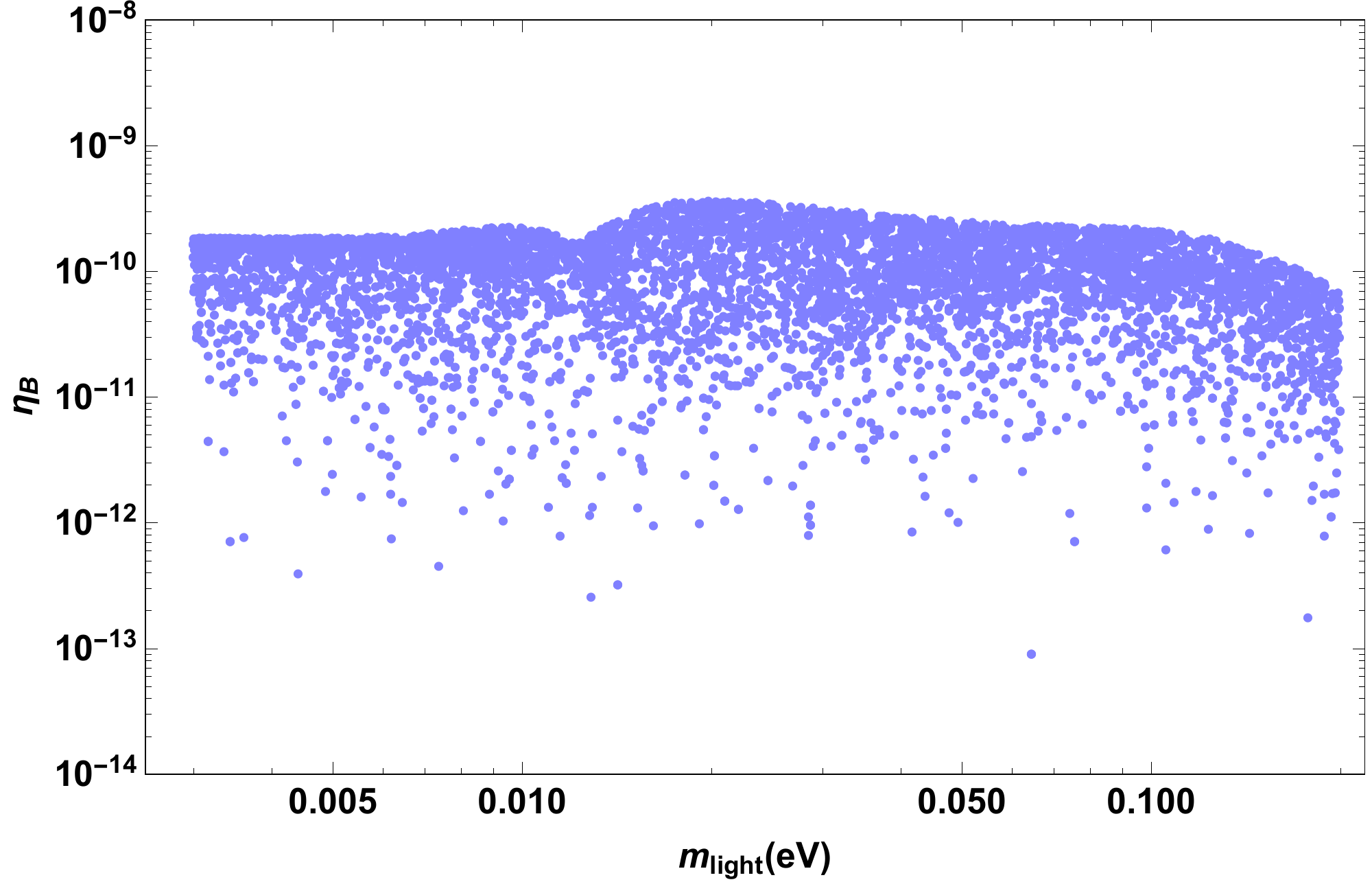}}
\hfil
\subfloat[$\alpha_{21}$ {\it vs} $\ml$ for the NO case]
{\includegraphics[width=0.45\textwidth]{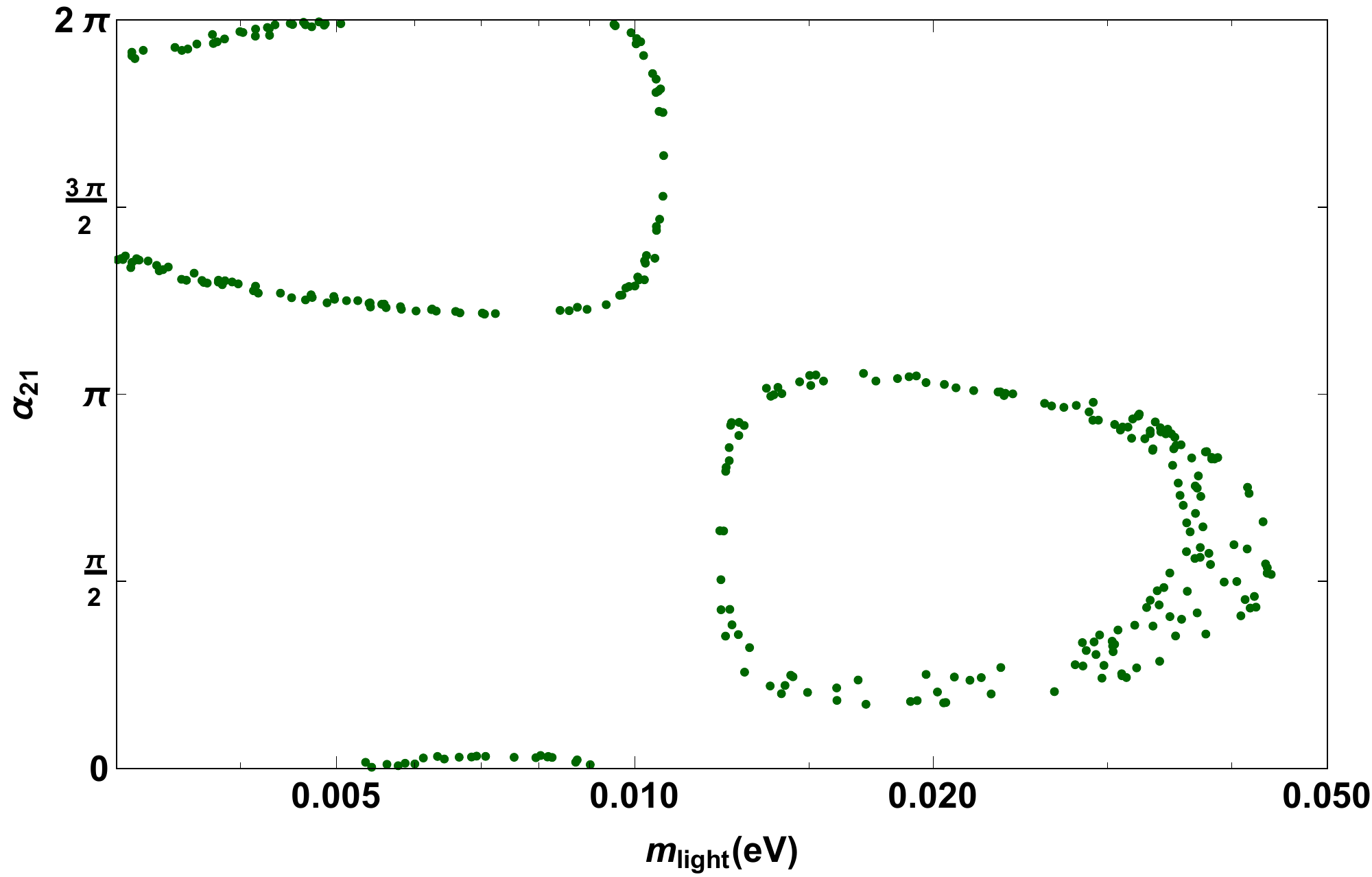}}
\caption{\footnotesize\em $\eta_B$ as a function of the lightest neutrino mass for the NO on the top and IO in the middle. In black the points where $\eta_B$ falls inside its experimental determination at $3\sigma$ error. The correlation between $\alpha_{21}$ and $\ml$ in the bottom for the NO case only: the points corresponds to the black ones in the first plot with $\eta_B$ inside its experimental value. Charged lepton masses and neutrino oscillation parameters have been taken at their central value as in Tab.~\ref{TableOscFit}, $0.01\lesssim z<20$, $c_1=c_2=0.01$ and the Majorana CP phases randomly vary in their dominium.}
\label{fig:EtaBLow}
\end{figure}

Lowering $T_\text{max}$ implies that the heaviest sterile neutrinos may not be thermally produced, preventing in this way their contributions to the final value of $\eta_B$. Fig.~\ref{fig:EtaBLow} shows the results for $T_\text{max}=10^{15}\GeV$. Comparing these plots with those in Fig.~\ref{fig:EtaB}, the NO case is essentially unaffected by this change, as also confirmed by the correlation plot showing the behaviour of the Majorana phase $\alpha_{21}$ {\it vs} $\ml$ when compared with the equivalent plot in Fig.~\ref{fig:CorrelationMajorana}(a). The IO case presents a sustancial difference, as $\eta_B$ does not reach the experimental band for small values of $\ml$.

\providecommand{\href}[2]{#2}\begingroup\raggedright\endgroup

\end{document}